\documentclass[12pt]{article}
\usepackage[latin1]{inputenc}
\usepackage[all]{xy}
\usepackage{graphicx}
\usepackage{latexsym}
\usepackage{color}
\usepackage{epsfig}
\usepackage{amsmath}
\usepackage{amssymb}
\usepackage{mathabx}
\usepackage{MnSymbol}

\def\l[{\left[}
\def\r]{\right]}

\newcommand{\ti}{\;\;\makebox[0pt]{$\top$}\makebox[0pt]{$\cap$}\;\;}
\def\ba{\begin{array}}
\def\ea{\end{array}}
\def\beq{\begin{equation}}
\def\eeq{\end{equation}}
\def\bea{\begin{eqnarray}}
\def\eea{\end{eqnarray}}

\definecolor{brique}{rgb}{.9,.2,0}
\definecolor{blvert}{rgb}{0,.8,.85}
\definecolor{vertcl}{rgb}{0,1,.7}
\newcommand\vertcl[1]{\textcolor{vertcl}{#1}}
\newcommand\blvert[1]{\textcolor{blvert}{#1}}
\newcommand\brique[1]{\textcolor{brique}{#1}}
\def\lapth{
\begin{picture}(136,70)(0,-15)\thicklines
\put(0,0){\vertcl{\rule{20pt}{4pt}}}
\put(19,1){\vertcl{\line(1,3){23}}}
\put(20,1){\vertcl{\line(1,3){23}}}
\put(21,1){\vertcl{\line(1,3){23}}}
\put(22,1){\vertcl{\line(1,3){23}}}
\put(45,70){\vertcl{\line(1,-3){23}}}
\put(44,70){\vertcl{\line(1,-3){23}}}
\put(43,70){\vertcl{\line(1,-3){23}}}
\put(42,70){\vertcl{\line(1,-3){23}}}
\put(2,24){\vertcl{\rule{120pt}{4pt}}}
\put(65,0){\vertcl{\rule{60pt}{4pt}}}
\put(5,37){\Huge{\brique{\textbf{L}}}}
\put(62,37){\Huge{\brique{\textbf{PTh}}}}
\put(12,-8){\blvert{\rule{92pt}{3.5pt}}}
\put(24,-15){\blvert{\rule{57pt}{3.5pt}}}
\put(36,-22){\blvert{\rule{30pt}{3.5pt}}}
\end{picture}
\raisebox{35pt}{
\begin{minipage}{320pt}\begin{center}
\textbf{Laboratoire d'Annecy-leVieux de Physique Th\'eorique}\\[4ex]
website: \texttt{http://lapth.in2p3.fr/}
\end{center}
\end{minipage}}\\
\vspace{10pt}\quad \hrulefill\\
\vspace{10pt}}

\begin{document}

\pagestyle{empty}
\setcounter{page}{0}
\hspace{-1cm}\lapth

\begin{center}
{\Large {\bf Higher dimensional abelian Chern-Simons theories and their link invariants}}%
\\[1.5cm]

{\large L. Gallot, E. Pilon, F. Thuillier}

\end{center}

\vskip 0.7 truecm

{\it LAPTH, Universit\'e de Savoie, CNRS}

{\it 9, Chemin de Bellevue, BP 110, F-74941
Annecy-le-Vieux cedex, France}.

\vskip 0.7 truecm

\texttt{gallot@lapp.in2p3.fr, pilon@lapp.in2p3.fr, thuillie@lapp.in2p3.fr }

\vspace{20mm}

\centerline{{\bf Abstract}}

The role played by Deligne-Beilinson cohomology in establishing the relation between
Chern-Simons theory and link invariants in dimensions higher than three is investigated.
Deligne-Beilinson cohomology classes provide a natural abelian Chern-Simons action, non
trivial only in dimensions $4l+3$, whose parameter $k$ is quantized. The generalized Wilson $(2l+1)$-loops are observables of the theory and their charges are quantized. The Chern-Simons action is then used to compute invariants for links of $(2l+1)$-loops,
first on closed $(4l+3)$-manifolds through a novel geometric computation, then on
$\mathbb{R}^{4l+3}$ through an unconventional field theoretic computation.


\vspace{5mm}

\indent

\vfill

\rightline{LAPTH-030/12}

\newpage
\pagestyle{plain} \renewcommand{\thefootnote}{\arabic{footnote}}

\section{Introduction}

The role that Deligne-Beilinson cohomology \cite{De,Be,EV,Ja,Br,MP,BGST} plays in establishing the
relation between Chern-Simons Quantum Field Theory and link invariants
\cite{Sc,Ha,Wi,Jo,RT,Ro,Mo,GMM,Gu}, in the abelian case, has been stressed out in a series of papers
\cite{GT,Th}. We will here complete these works by showing how higher dimensional Deligne-Beilinson
(DB) cohomology classes, and their DB-products, provide a natural generalisation of the
Chern-Simons action, and how they can be used to compute invariants for higher dimensional links \cite{Ro,GCSS}.
We will produce a novel, geometric computation for closed $(4l+3)$-manifolds. We will then
compare it to a field theoretic computation made on $\mathbb{R}^{4l+3}$.

In section 2, we recall some basic facts concerning Deligne-Beilinson cohomology and how it relates
to the functional measure based on the abelian Chern-Simons action. In section 3, we present a
natural candidate for the generalized CS action. In section 4, we deal with generalized abelian
loops and their expectation values for closed $(4l+3)$-manifolds within the DB approach.
We further illustrate it with two specific examples. Section 5 is devoted to
a quite unusual field theoretic computation of these expectation values in the $\mathbb{R}^{4l+3}$
case, and the extension of this type of computation to $S^{4l+3}$ is sketched. In Appendix, a
geometrical interpretation of the higher dimensional linking number relating it to the notions of
solid angle and zodiacus is presented following the original ideas of Gauss \cite{GE}.

$\\$

\noindent
Here are the main results elaborated in this article:

\begin{enumerate}

\item

The abelian Chern-Simons generalised action is non trivial only in dimension $4l+3$, and its
level parameter $k$ has to be quantized;

\item
The generalised Wilson $(2l+1)$-loops are observables of the theory and their charges are
quantized.

\item
In the geometric DB approach provided by functional integration over the space
$[H_D^{2l+1} \left({M,{\mathbb Z}} \right)]^* \supset H_D^{2l+1} \left({M,{\mathbb Z}} \right)$, the $2k$-nilpotency property holds and the observables are given by (self-)linking numbers under the so-called zero-regularization choice ({\em i.e.} framing). Furthermore only homology is involved in abelian Chern-Simons theories and only homologically trivial links (modulo 2k) give non vanishing expectation values.

\item
A field theoretic computation in $\mathbb{R}^{4l+3}$ can be handled in a non perturbative way, yet
it still misses quantization of the level and charges. Once the latter are imposed by hand the
result reproduces the one from the DB approach.

\end{enumerate}

$\\$

\section{Basic facts about Deligne-Beilinson cohomology}

Without recalling the whole theory let us remind the basic facts about DB-cohomology useful in this
paper.

\subsection{Definition via exact sequences}

If $M$ is a closed (\textit{i.e.} compact and without boundary) $n$-dimensional smooth manifold, the p-th DB cohomology group
of $M$, denoted $H_D^p \left({M,{\mathbb Z}} \right)$  ($p\leq dim M = n$), is canonically embedded
into the following equivalent exact sequences \cite{Br,HLZ}:
\bea
0 \buildrel \over \longrightarrow {\Omega^p\left( M \right)} \mathord{\left/
{\vphantom {{\Omega^p\left( M \right)} {\Omega_{\mathbb Z}^p \left( M
\right)}}} \right. \kern-\nulldelimiterspace} {\Omega_{\mathbb Z}^p \left( M
\right)} \buildrel \over \longrightarrow &H_D^p \left( {M,{\mathbb Z}}
\right)& \buildrel \over \longrightarrow \check{H}^{p+1}\left( {M,{\mathbb Z}}
\right)\buildrel \over \longrightarrow 0 \, ,
\label{1bis}
\\
0\buildrel \over \longrightarrow \check{H}^{p}\left( {M,{\mathbb R
}/{\mathbb Z}}\right) \buildrel
\over \longrightarrow &H_D^p \left( {M,{\mathbb Z}}
\right)& \buildrel \over \longrightarrow \Omega _{\mathbb Z}^{p+1} \left( M \right)\buildrel
\over \longrightarrow 0 \, ,
\label{1bisbis-unused}
\eea
where $\Omega^p\left( M \right)$ is the space of smooth $p$-forms on $M$,
$\Omega_{\mathbb Z}^p \left( M \right)$ the space of smooth closed p-forms with
integral periods on $M$, $\check{H}^{p+1}\left( {M,{\mathbb Z}} \right)$ is the
$(p+1)$-th integral $\check{\textrm{C}}$ech cohomology group of $M$, and
$\check{H}^{1}\left( {M,{\mathbb R}/{\mathbb Z}}\right)$ is the $p$-th ${\mathbb R
}/{\mathbb Z}$-valued $\check{\textrm{C}}$ech cohomology group of $M$. These exact sequences also occur in the context of Cheeger-Simons differential characters \cite{CS,Ko} or Harvey-Lawson sparks \cite{HLZ}.

Thanks to exact sequences (\ref{1bis}) one can interpret $H_D^p \left( {M,{\mathbb Z}}
\right)$ as an affine bundle over $\check{H}^{p+1}\left( {M,{\mathbb Z}}
\right)$ (resp. $\Omega _{\mathbb Z}^{p+1} \left( M \right)$) with structure group
${\Omega^p\left( M \right)} \mathord{\left/
{\vphantom {{\Omega^p\left( M \right)} {\Omega_{\mathbb Z}^p \left( M
\right)}}} \right. \kern-\nulldelimiterspace} {\Omega_{\mathbb Z}^p \left( M
\right)}$ (resp. $\check{H}^{p}\left( {M,{\mathbb R
}/{\mathbb Z}}\right)$).
Note that in the former case $\Omega_{\mathbb Z}^p \left( M \right)$ plays the role of a gauge
group, which is much bigger (in general) than the usual group of exact forms. An element of
$H_D^p \left( {M,{\mathbb Z}}\right)$ will be generically written $\omega^{[p]}$.

Let us pick up a normalized volume form on $M$, \textit{i.e.} a $n$-form $\mu$ such that $\int_M
\mu = 1$. For dimensional reasons any $n$-form on $M$ is closed, hence for any $n$-form $\omega$
on $M$ there exists a $(n-1)$-form $\nu$ such that $\omega = \tau \mu + d \nu$, with $\tau =
\int_M \omega \in \mathbb{R}$. Furthermore, if $\omega$ has integral periods, then $\tau \in
\mathbb{Z}$, since $d \nu$ is a closed $n$-form with zero periods ($\int_M d \nu = 0$ since $M$
has no boundary). This proves that any element of ${\Omega^n(M)} / {\Omega_{\mathbb{Z}}^{n}(M)}$
can be written as $\theta \mu$, with $\theta \in \mathbb{R}/\mathbb{Z}$. Finally, integrating
$\theta \mu$ over $M$ makes the construction independent of $\mu$ and proves that ${\Omega^n(M)}
/ {\Omega_{\mathbb{Z}}^{n}(M)} \simeq \mathbb{R}/\mathbb{Z}$ (equivalently one can pick up
another normalized volume form and see that it will give the same $\theta$, and finally pick any
volume form and prove the same). Still for dimensional reasons, $\check{H}^{n+1}(M,\mathbb{Z}) =
0$, so we conclude that $H_D^n \left( {M,{\mathbb Z}} \right) \simeq \mathbb{R}/\mathbb{Z}$.

For later convenience, let us consider two special cases. First, when $M=S^{4l+3}$ and $p=2l+1$, we
have $\check{H}^{2l+1}\left({M,{\mathbb R}/{\mathbb Z}}\right) = 0 =
\check{H}^{2l+2}\left({M,{\mathbb R}/{\mathbb Z}}\right)$, then sequence (\ref{1bis})
reduces to:
\bea
\label{example1}
0 \buildrel \over \longrightarrow {\Omega^{2l+1}(M)} \mathord{\left/
{\vphantom {{\Omega^p\left( M \right)} {\Omega_{\mathbb Z}^{2l+1}(M)}}} \right.
\kern-\nulldelimiterspace} {d \Omega^{2l}(M)} \buildrel \over \longrightarrow &H_D^{2l+1}
\left( {M,{\mathbb Z}} \right)& \buildrel \over \longrightarrow 0 \, .
\eea

\noindent
Hence $H_D^{2l+1} \left( {M,{\mathbb Z}}\right)$ is isomorphic to the quotient
space ${\Omega^{2l+1}(M)} / {d \Omega^{2l}(M)}$, the gauge group reducing to the trivial group
$d \Omega^{2l}(M)$. Although this is a quite trivial case, it is very close to the one of the
field theoretic approach.

The second example is provided by $M=S^{2l+1} \times S^{2l+2}$, still with $p=2l+1$. Since
$\check{H}^{2l+1}\left({M,{\mathbb R}/{\mathbb Z}}\right) = \mathbb Z =
\check{H}^{2l+2}\left({M,{\mathbb R}/{\mathbb Z}}\right)$, sequence (\ref{1bis}) reads:
\bea
\label{example2}
0 \buildrel \over \longrightarrow {\Omega^{2l+1}\left( M \right)} \mathord{\left/
{\vphantom {{\Omega^{2l+1}\left( M \right)} {\Omega_{\mathbb Z}^{2l+1} \left( M
\right)}}} \right. \kern-\nulldelimiterspace} {\Omega_{\mathbb Z}^{2l+1} \left( M
\right)} \buildrel \over \longrightarrow &H_D^{2l+1} \left( {M,{\mathbb Z}}
\right)& \buildrel \over \longrightarrow \check{H}^{2l+2}\left( {M,{\mathbb Z}}
\right) = \mathbb{Z} \buildrel \over \longrightarrow 0 \, .
\eea
The DB $\mathbb{Z}$-module $H_D^{2l+1} ({M,{\mathbb Z}})$ is then a non trivial affine bundle
over $\mathbb{Z}$, the gauge group $\Omega_{\mathbb Z}^{2l+1}(M)$ being also now non trivial.

\subsection{Pontrjagin dual of DB-spaces}

Due to the form of the exact sequences (\ref{1bis}), one can consider dual sequences not with
respect to $\mathbb{R}$ but to $\mathbb{R} / \mathbb{Z}$. This gives rise to the Pontrjagin dual
space of $H_D^{p} \left( {M,{\mathbb Z}}\right)$: $H_D^p \left( {M,{\mathbb Z}} \right)^{\ast}
\equiv Hom(H_D^p \left( {M,{\mathbb Z}} \right),S^1)$. In particular, $H_D^p \left( {M,{\mathbb Z}}
\right)^{\ast}$ belongs itself to an exact sequence (dualizing (\ref{1bisbis-unused}) in
$\mathbb{R} / \mathbb{Z}$):
\begin{equation}
\label{8bis}
0\buildrel \over \longrightarrow Hom\left( \Omega _{\mathbb Z}^{p+1}
\left( M \right),{{\mathbb R}/{\mathbb Z}} \right) \buildrel
 \over \longrightarrow H_D^p \left( {M,{\mathbb Z}}
\right)^{\ast}\buildrel \over \longrightarrow \check{H}^{n-p-1}\left( {M,{\mathbb Z}}
\right) \buildrel \over \longrightarrow 0 \, ,
\end{equation}
This identifies $H_D^p \left( {M,{\mathbb Z}} \right)^{\ast}$ as an affine bundle over the same
base, $\check{H}^{n-p-1}\left( {M,{\mathbb Z}} \right)$, than $H_D^{n-p-1} \left( {M,{\mathbb Z}}
\right)$. Of course there is a second exact sequence we could obtain from dualizing (\ref{1bis}).

Thanks to integration over integral cycles on $M$, the quotient
${\Omega^{n-p-1}\left( M \right)} \mathord{\left/{\vphantom {{\Omega^{n-p-1}\left( M \right)}
{\Omega_{\mathbb Z}^{n-p-1} \left( M \right)}}} \right. \kern-\nulldelimiterspace}
{\Omega_{\mathbb Z}^{n-p-1} \left( M \right)}$ can be canonically embedded into
$Hom\left( \Omega _{\mathbb Z}^{p+1} \left( M \right),{{\mathbb R}/{\mathbb Z}} \right)$.
We have also noticed that
$H_D^{n}(M,\mathbb{Z})  \simeq \mathbb{R}/\mathbb{Z}$. This suggests that
$H_D^{n-p-1} \left( {M,{\mathbb Z}} \right)$ might be canonically identified as a subset of
$H_D^p \left( {M,{\mathbb Z}} \right)^{\ast}$, just as continuous functions can be seen as
(regular) distributions. The notion of integration of DB-classes over cycles is needed to confirm
this.

\subsection{Integration of DB-classes over integral cycles}

There is a canonical pairing between DB-class and cycles on $M$ provided by integration of the
later over the former:
\begin{equation}
\label{6bis}
\oint : H_D^p \left( {M,{\mathbb Z}} \right) \times Z_p \left( {M}
\right)\longrightarrow {{\mathbb R}/{\mathbb Z}} \, ,
\end{equation}
where $Z_p(M)$ denotes the space of  integral $p$-cycles on $M$. Let us stress that these integrals
take their values in $\mathbb{R}/\mathbb{Z} \simeq S^1$, not $\mathbb{R}$.

Since $M$ itself is a cycle, one can integrate any DB-class
$\omega^{[n]} \in H_D^n \left( {M,{\mathbb Z}} \right)$ over $M$. This confirms that
$H_D^{n}(M,\mathbb{Z}) \simeq \mathbb{R}/\mathbb{Z}$ and proves that
$H_D^{n-p-1} \left( {M,{\mathbb Z}} \right)$ can be canonically identified as a subset of
$H_D^p \left( {M,{\mathbb Z}} \right)^{\ast}$.

Incidentally, integration also shows that $Z_p(M)$ is canonically embedded into
$H_D^p \left( {M,{\mathbb Z}} \right)^{\ast}$ - which can be expressed \cite{HLZ} by
saying that $p$-cycles live in the topological boundary of
$H_D^p \left( {M,{\mathbb Z}} \right)^{\ast}$. Hence:
\begin{equation}
\label{10bis}
H_D^{n-p-1} \left( {M,{\mathbb Z}}
\right) \times Z_p(M) \subset H_D^p \left( {M,{\mathbb Z}}
\right)^{\ast} \, ,
\end{equation}
where $\subset$ has to be understood as the above canonical embeddings.

\vspace{5mm}

\noindent \textbf{Property 1} \textit{As in the three dimensional case, abelian holonomies
defined by:}
\begin{equation}
\label{9ter}
\exp \left\{2i \pi \oint_z \omega^{[p]} \right\} ,
\end{equation}
\textit{are observables of the generalized abelian Chern-Simons theories.}

\subsection{DB-product and cycle map}

There is a natural bilinear product, referred here as the DB-product:
\begin{equation}
\label{2bis}
\ast_D : H_D^p \left( {M,{\mathbb Z}} \right) \times H_D^q \left( {M,{\mathbb Z}}
\right)\longrightarrow H_D^{p+q+1} \left( {M,{\mathbb Z}}
\right) \, ,
\end{equation}
which is graded according to:
\begin{equation}
\label{3bis}
\omega_1^{[p]} \ast_D \omega_2^{[q]} = (-1)^{(p+1)(q+1)} \omega_2^{[p]} \ast_D \omega_1^{[q]} \, .
\end{equation}

\noindent From our previous remarks, one straightforwardly verifies:
\begin{equation}
\ast_D : H_D^p(M,\mathbb{Z}) \times
H_D^{n-p-1}(M,\mathbb{Z}) \longrightarrow H_D^{n}(M,\mathbb{Z})  \simeq \mathbb{R}/\mathbb{Z}
\end{equation}
The ``DB-square" operation satisfies the graded commutation property:
\begin{equation}
\label{4bis}
\omega^{[p]} \ast_D \omega^{[p]} = (-1)^{(p+1)(p+1)} \omega^{[p]} \ast_D \omega^{[p]} \, .
\end{equation}
which implies in particular:
\begin{equation}
\label{5bis}
\omega^{[2l]} \ast_D \omega^{[2l]} = 0 \, ,
\end{equation}
for any $\omega^{[2l]} \in H_D^{2l}(M,\mathbb{Z})$.

The DB-classes introduced above are smooth ones. They can be extended to distributional DB-classes.
relying on Pontrjagin duality. Setting $H_D^{-1}(M,\mathbb{Z}) \equiv \mathbb{Z}$, one extends the
previous DB-product to a pairing of $H_D^p(M,\mathbb{Z})$ and $H_D^q(M,\mathbb{Z})^{\ast}$ into
$H_D^{(q-p-1)}(M,\mathbb{Z})^{\ast} \supset H_D^{(n-q+p+1)}(M,\mathbb{Z})$ ($q \geq p$). Note that
$H_D^{-1}(M,\mathbb{Z})^{\ast} = \mathbb{R} / \mathbb{Z} = H_D^{n}(M,\mathbb{Z})$ hence $\ast_D
: H_D^p(M,\mathbb{Z}) \times H_D^p(M,\mathbb{Z})^{\ast} \rightarrow H_D^{-1}(M,\mathbb{Z})^{\ast} =
\mathbb{R} / \mathbb{Z}$ as expected. This is similar to the usual theory of de Rham currents.

We end this subsection with the following important result shown in \cite{BGST}: to any
$p$-cycle $z$ on $M$ one can associate a canonical distributional DB-class
$\eta_z \in H_D^p(M,\mathbb{Z})^{\ast}$ such that:
\begin{equation}
\label{9bis}
\oint_z \omega^{[p]}  \mathop  =
\int_M \omega^{[p]} \ast_D  \eta_z  \, ,
\end{equation}
for any $\omega^{[p]} \in H_D^p \left( {M,{\mathbb Z}}\right)$. Such distributional DB-classes thus
appear as elements of $H_D^p \left( {M,{\mathbb Z}} \right)^{\ast}$. This is just another way to
see the inclusion  $Z_p(M) \subset H_D^p \left( {M,{\mathbb Z}} \right)^{\ast}$. In the particular
case where the $p$-cycle is a boundary, $z = bc$, the associated DB-class $\eta_z^{[n-p-1]}$
reduces to the de Rham current of the integral $(p+1)$-chain $c$. See \cite{BGST} for details.

\section{Generalized Chern-Simons action, Chern-Simons functional measure, observables and
framing}\label{section3}

\subsection{Generalized Chern-Simons action}\label{subsection3.1}

It is standard from a physicist point of view to present the abelian Chern-Simons (CS) lagrangian
on $\mathbb{R}^3$ as :
\begin{equation}
\label{11bis}
cs_1(A) \equiv A \wedge dA  \, ,
\end{equation}
or, using the CS action:
\begin{equation}
\label{12bis}
CS_1(A) = 2i \pi \int_{\mathbb{R}^3} A \wedge dA  \, ,
\end{equation}
where $A$ is a  $U(1)$-connection on some principal $U(1)$-bundle $P$ over $\mathbb{R}^3$.  A natural
generalization for $\mathbb{R}^{4l+3}$ would be to replace $A$ in eqn. \ref{11bis} by a
$(2l+1)$-form.
This is what will be done in section 5 when dealing with the field theoretic formulation.

However $U(1)$-connections on $M$ are actually not $1$-forms for compactclosed $3$-manifolds $M$.
Hence, as explained in \cite{GT,Th}, we rather have to use
DB-classes to write the lagrangian (\ref{11bis}), and hence the action (\ref{12bis}).
Let us recall that $H_D^1 \left(
{M,{\mathbb Z}}\right)$ canonically identifies with the set of classes of $U(1)$-isomorphic
principal $U(1)$-bundles with connection over $M$. Hence we must replace eqn. (\ref{12bis}) by
\begin{equation}
\label{13bis}
CS_1(A) = 2i \pi \int_M A \ast_D A  \, ,
\end{equation}
where $A$ has now to be understood as a DB class.

For a level $k$ CS theory we set:
\begin{equation}
\label{14bis}
CS_k(A) = 2i \pi k \int_M A \ast_D A  \, .
\end{equation}

We can extend the definition of the action (\ref{14bis}) to any closed smooth $n$-dimensional manifold $M$ as:
\begin{equation}
\label{15bis}
CS_k(\omega^{[p]}) = 2i \pi k \int_M \omega^{[p]} \ast_D \omega^{[p]}  \, .
\end{equation}
This will be our definition of the $n$-dimensional Chern-Simons theory of level $k$ on $M$.
Since integrals take values in  $\mathbb{R}/\mathbb{Z}$ this quantity is well defined provided
\begin{equation}
\label{17bis}
k \in \mathbb{Z} \, ,
\end{equation}
which is the announced \textbf{quantization of the level parameter}.

We now consider the ``quantum weight":
\begin{equation}
\label{17bis}
\exp \left\{ CS_k(\omega^{[p]}) \right\} =
\exp \left\{ 2i \pi k \int_M \omega^{[p]} \ast_D \omega^{[p]} \right\}  \, .
\end{equation}
When $p=2l$ the graded commutation property (\ref{4bis}) leads to:
\begin{equation}
\label{18bis}
\exp \left\{ CS_k(\omega^{[2l]}) \right\} =
\exp \left\{ 2i \pi k \int_M \omega^{[2l]} \ast_D \omega^{[2l]} \right\} = 1 \, .
\end{equation}
thereby providing a trivial functional measure. Consequently, the
non-trivial cases only occur when $p=2l+1$  which implies that $n = 2p+1 =\textbf{4l+3}$. In
particular, if $M$ is a sphere, the only non trivial abelian Chern-Simons theories will occur for
\begin{equation}
\label{19bis}
S^3 \, , \, S^7 \, , \, S^{11} \,  ... \, .
\end{equation}
Note that this is namely the set of spheres for which Hopf invariants are non-trivial, hence
linking numbers are non trivial as well \cite{BT82}. Furthermore, this expression for the CS action holds true for closed manifolds with torsion.

In summary:

\vspace{5mm}

\noindent \textbf{Property 2}
\textit{The non trivial generalized abelian Chern-Simon lagrangian of level} $k$  \textit{is defined by the DB square product of (2l+1) dimensional DB classes on a (4l+3)-dimensional closed
manifold, with} $k$ \textit{an integer}.

\vspace{5mm}

For a $(4l+3)$-dimensional manifold and its $(2l+1)$-loops, the inclusions stressed out after
(\ref{8bis}) and in (\ref{10bis}) give:
\bea
\label{21ter}
H_D^{2l+1} \left( {M,{\mathbb Z}}
\right) &\subset& H_D^{2l+1} \left( {M,{\mathbb Z}}
\right)^{\ast} \,  , \\  \, {\Omega^{2l+1}\left( M \right)} \mathord{\left/
{\vphantom {{\Omega^{2l+1}\left( M \right)} {\Omega_{\mathbb Z}^{2l+1} \left( M
\right)}}} \right. \kern-\nulldelimiterspace} {\Omega_{\mathbb Z}^{2l+1} \left( M
\right)} &\subset& Hom\left( \Omega _{\mathbb Z}^{2l+2} \left( M \right),{{\mathbb R}/{\mathbb Z}} \right) \, \nonumber .
\eea

We will assume that the space of quantum fields of a generalized abelian Chern-Simons theory in
$(4l+3)$ dimensions is a subset of $H_D^{2l+1} \left( {M,{\mathbb Z}} \right)^{\ast}$ which contains
$H_D^{2l+1} \left( {M,{\mathbb Z}} \right) \times Z_{2l+1}(M)$.

\subsection{Chern-Simons functional measure and zero mode property}\label{subsection3.2}

The generalized Chern-Simons ``gaussian" functional measure for a $(4l+3)$-manifold takes
the form:

\begin{equation}
\label{20bis}
d\mu_k(\omega) \equiv D \omega \exp \left\{ CS_k(\omega) \right\}  \, .
\end{equation}

Since we wish to use this measure to compute observables and identify them with $(2l+1)$-links
invariants, let us have a closer look at it. First, $d\mu_k(\omega)$ is supposed to be a measure on
$H_D^{2l+1} \left( {M,{\mathbb Z}} \right)$ or rather on (some subset of)
$H_D^{2l+1} \left( {M,{\mathbb Z}} \right)^{\ast}$, its ``quantum" version. Of course, and as
usual for infinite dimensional spaces, the measure (\ref{20bis}) is totally formal on both spaces: as
a Lebesgue measure over $H_D^{2l+1} \left( {M,{\mathbb Z}} \right)$, $D \omega$ is zero, and so is
(\ref{20bis}); considering globally on $d\mu_k(\omega)$
$H_D^{2l+1} \left( {M,{\mathbb Z}} \right)^{\ast}$, we should need to regularize products of
distributional DB classes appearing in the gaussian part of the measure -something common
in Quantum Field Theory. In fact, we will only need the fundamental Cameron-Martin like
property for the measure (\ref{20bis}), that is to say:
\begin{equation}
\label{21quart}
d\mu_k(\omega + \zeta) = d\mu_k(\omega)
\exp\left\{ 4i \pi k \int_{M} \omega \ast_D \zeta  \right\}
\exp\left\{ 2i \pi k \int_{M} \zeta \ast_D \zeta  \right\} \, ,
\end{equation}
for any given $\zeta \in H_D^{2l+1} \left( {M,{\mathbb Z}} \right)$. Note that this property is
similar to the one of a finite-dimensional gaussian measure which relies on the translational
invariance of the Lebesgue measure. In other words, we have to assume that the ``existing
measure" on the functional space has property (\ref{21quart}) which holds true for (\ref{20bis})
seen has a measure on any finite dimensional subset of $H_D^{2l+1} \left( {M,{\mathbb Z}} \right)$.

Let us consider a $(2l+2)$-cycle $\Sigma$, whose integration $(2l+1)$-current in $M$ is denoted
$\beta_{\Sigma}$. While this current canonically represents the zero class in
$H_D^{2l+1} \left( {M,{\mathbb Z}} \right)$, in general the current $\frac{\beta_{\Sigma}}{2k}$
does not. From property (\ref{21quart}), and identically denoting currents and the DB classes which they
represent, we deduce:
\begin{equation}
\label{21invariant}
d\mu_k(\omega + \frac{\beta_{\Sigma}}{2k}) = d\mu_k(\omega)
\exp\left\{ 4i \pi k \int_{M} \omega \ast_D \frac{\beta_{\Sigma}}{2k}  \right\}
\exp\left\{ 2i \pi k \int_{M} \frac{\beta_{\Sigma}}{2k} \ast_D \frac{\beta_{\Sigma}}{2k}  \right\}
\, .
\end{equation}
In contrast with the identity
\begin{equation}
\label{21inv2}
\exp\left\{ 2i \pi k \int_{M} \frac{\beta_{\Sigma}}{2k} \ast_D \frac{\beta_{\Sigma}}{2k}  \right\}
=
\exp\left\{ \frac{2i \pi}{4k} \int_{M} \beta_{\Sigma} \wedge d \beta_{\Sigma}  \right\} = 1 \, ,
\end{equation}
trivial since $d \beta_{\Sigma} = 0$, the following one:
\begin{equation}
\label{21inv1}
\exp\left\{ 4i \pi k \int_{M} \omega \ast_D \frac{\beta_{\Sigma}}{2k}  \right\}
= \exp\left\{ 2i \pi \int_{M} \omega \ast_D \beta_{\Sigma}  \right\} = 1 \, ,
\end{equation}
\noindent
deserves some justification. The factor $4 i \pi k = 2k \cdot (2 i \pi)$ in eqn. (\ref{21inv1}) is of
pivotal importance. Indeed, $\omega \ast_D \beta_{\Sigma} / 2k$ is not the zero class,
whereas $2k (\omega \ast_D \beta_{\Sigma} / 2k) = \omega \ast_D \beta_{\Sigma}$ is, as
$\beta_{\Sigma}$ is trivial. Note that $\beta_{\Sigma} / 2k$ is not an integer current, and that a
DB class $\omega$ is not the restriction of a current in general (see for instance \cite{BGST}).
Of course, one should be careful when dealing with the product of currents
$\beta_{\Sigma} \wedge d \beta_{\Sigma}$. However one can always smooth $\beta_{\Sigma}$ around
$\Sigma$ (\textit{i.e.} use a Poincaré representative with
support as close to $\Sigma$ as necessary) in order to consistently regularize
$\beta_{\Sigma} \wedge d \beta_{\Sigma}$ to the zero DB class.
More generally, for any integer $m$,
\begin{equation}\label{21invariant-m}
d\mu_k(\omega + m \, \frac{\beta_{\Sigma}}{2k}) = d\mu_k(\omega)
\end{equation}
which provides the generalization of Property 4 of \cite{GT}:

\vspace{5mm}

\noindent \textbf{Property 3}
\textit{The functional measure $d\mu_k(\omega)$ is invariant under translations by
$m\frac{\beta_{\Sigma}}{2k}$, where $\beta_{\Sigma}$ is the integration current of a $(2l+2)$-cycle
$\Sigma$ and $m$ an integer.}

\vspace{5mm}

When $\Sigma$ is homologically trivial ($\Sigma = b \cal{V}$) then $\beta_{\Sigma} = d
\chi_{\cal{V}}$, and therefore $\frac{\beta_{\Sigma}}{2k} = d (\frac{\chi_{\cal{V}}}{2k})$ . In this
case the DB-class of $\frac{\beta_{\Sigma}}{2k}$ is also zero. This happens for any $\Sigma$ when the
$(2l+2)$th homology group of $M$ is trivial. Conversely, as we shall see in the next section, when
$M$ has a non trivial $(2l+2)$-th homology group, Property 3 will provide a treatment of the so-called "zero modes", thus leading to the important result of this paper concerning the
vanishing of links invariants.

\subsection{Observables and Framing}\label{subsection3.3}

Following Property 1, let us consider an observable of our level $k$ generalized CS theory:
\begin{equation}
\label{observables}
\exp \left \{ 2i \pi \oint_z \omega \right \} =
\exp \left \{ 2i \pi \int_M \omega \ast_D  \eta_z \right \} \, .
\end{equation}
Let us remind that a $(2l+1)$-loop is meant to be a continuous mapping
$\gamma : \Sigma_{2l+1} \rightarrow M$, where $\Sigma_{2l+1}$ is a closed $(2l+1)$-dimensional manifold. It is always possible to identify such a loop with a $(2l+1)$-cycle in $M$.
Furthermore, if the mapping is an embedding ({\em i.e.} the image $\gamma(\Sigma_{2l+1})$ is
isomorphic to $\Sigma_{2l+1}$) $\gamma$ is said to be a \textbf{fundamental loop}.
Then, seen as a cycle, any $(2l+1)$-loop in $M$ can be written as: $\gamma = q \gamma_0$, for some
fundamental loop $\gamma_0$ and $q \in \mathbb{Z}$.
Hence, the abelian Wilson line of the gauge field $\omega$ of degree $(2l+1)$ along a
$(2l+1)$-loop $\gamma = q \gamma_{0}$ in $M$ reads:

\begin{equation}
\label{Wilson}
W(\omega, \gamma) \equiv  \exp \left\{ 2i \pi \oint_{\gamma} \omega  \right\}
= \exp \left \{ 2i \pi q \int_{\gamma_0} \omega \right \} \, ,
\end{equation}
Conversely, the righthand side of this expression has a meaning if and only if $q$ is an integer.
This leads to:

\vspace{5mm}

\noindent
\textbf{Property 4} \textit{In the generalized CS theories, loops must have integer charges.}

\vspace{5mm}

The charge (or colour) of a loop $\gamma$ can be geometrically interpreted as the number of times the fundamental loop associated with $\gamma$ has been covered. When $\gamma$ is not homologically trivial, its charge canonically identifies with its homology class. The charge can also be seen has defining a representation for the $U(1)$ holonomy of a fundamental loop. This is also true for the level $k$ parameter which can be seen as a charge of $M$, or as a representation of the $U(1)$ 3-holonomy given by the Chern-Simons action.

\vspace{5mm}

If $\eta_{\gamma}$ and $\eta_0$ are the DB classes
($\in H_D^{2l+1} \left( {M,{\mathbb Z}} \right)^{\ast}$) associated with $\gamma$ and $\gamma_0$
respectively, then $\eta_{\gamma} = q \eta_0$. Hence we can alternatively write:
\begin{equation}
\label{Wilson2}
W(\omega, \gamma) = \exp \left \{ 2i \pi q \int_M \omega \ast_D \eta_0 \right \} \, .
\end{equation}

The expectation values of the Wilson lines are given by:
\begin{equation}
\label{WilsonMV}
< W(\omega, \gamma) >_{CS_k} =
Z_{k}^{-1} \int d\mu_k(\omega) \exp \left \{ 2i \pi q \int_M \omega \ast_D \eta_0 \right \} \, ,
\end{equation}
where $Z_{k}$ is the normalization factor such that $< W(\omega, \gamma \equiv 0) >_{CS_k} = 1$.

For a generic homological combination $\gamma = \sum_{i=1}^n q_i \gamma^0_i$ with
$q_i \in \mathbb{Z}$ and $\gamma^0_i$ fundamental, we get:
\begin{equation}
\label{WilsonGen}
W(\omega, \gamma) =
\exp \left \{ 2i \pi \sum_{i=1}^n q_i \int_{\gamma^0_i} \omega  \right \} \, ,
\end{equation}
or in term of the DB representatives $\eta^0_i$ of these $\gamma^0_i$:
\begin{equation}
\label{WilsonGen}
W(\omega, \gamma) =
\exp \left \{ 2i \pi \sum_{i=1}^n q_i \int_M \omega \ast_D \eta^0_i \right \} \, .
\end{equation}

Let us first exhibit the nilpotency property of the expectation values
\begin{equation}
\label{WilsonGenMV}
< W(\omega, \gamma) >_{CS_k} =
Z_{k}^{-1} \int d\mu_k(\omega) \exp \left \{ 2i \pi \sum_{i=1}^n q_i
\int_M \omega \ast_D \eta^0_i \right \} \, ,
\end{equation}
For the loop $2k\gamma_0$, where $\gamma_0$ is fundamental with DB representative $\eta_0$:
\begin{equation}
\label{period}
< W(\omega, 2k \gamma_0) >_{CS_k} =
Z_{k}^{-1} \int d\mu_k(\omega) \exp \left \{ 2i \pi (2k) \int_M \omega \ast_D \eta_0 \right \} \, .
\end{equation}
Performing the shift
\begin{equation}
\label{shift}
\omega \mapsto \omega + \eta_0 \, ,
\end{equation}
thanks to property (\ref{21quart}), we obtain:
\begin{equation}
\label{period2}
< W(\omega, 2k \gamma_0) >_{CS_k} =
Z_{k}^{-1} \int d\mu_k(\omega) \exp \left \{ - 2i \pi \int_M \eta_0 \ast_D \eta_0 \right \} \, .
\end{equation}
Such an expression is ill-defined since $\eta_0$ is distributional. If we decide to regularize the
quantities $\eta_0 \ast_D \eta_0$ into the zero DB class, which we refer to as the
\textbf{zero-regularization}, then:
\begin{equation}
\label{period2}
< W(\omega, 2k \gamma_0) >_{CS_k} = 1 = < W(\omega, \gamma \equiv 0) >_{CS_k} \, .
\end{equation}
This gives:

\vspace{5mm}

\noindent
\textbf{Property 5} \textit{The generalized CS theories satisfy the $2k$-nilpotency property}.

\vspace{5mm}

Zero-regularization calls for a comparison with framing.
If $\gamma_0$ is a boundary (\textit{i.e.} is homologically trivial), then
\begin{equation}
\label{regframing}
\int_M \eta_0 \ast_D \eta_0 \mathop  = \limits_\mathbb{Z} \int_M \chi_0 \wedge d \chi_0 \, ,
\end{equation}
where $\chi_0$ is the current of a chain whose $\gamma_0$ is the boundary, while $d \chi_0$ is
the de Rham current of $\gamma_0$. The symbol $\mathop = \limits_\mathbb{Z}$ in eqn. (\ref{regframing}) means
``equals modulo integers".
The framing procedure gives a meaning to the right hand side of eqn. (\ref{regframing}):
each framing choice assigns a well defined  {\em i.e.} homotopically invariant integer value to the
self-linking of $\gamma_{0}$. The difference between two choices of framing is an integer, which
coincides with taking $\eta_0 \ast_D \eta_0 = 0$.
However, when $\gamma_0$ is not a boundary the framing procedure is not a well-defined
regularization as it does not provide a definite homotopically invariant integer for the
self-linking number $\int_M \chi_0 \wedge d \chi_0$.
Notwithstanding property (\ref{period2}) still holds, the zero-regularization is thus coarser than framing yet more ``general".
Let us point out that $2k$-nilpotency\begin{footnote}{This was called colour periodicity
in \cite{GT}. Yet the name ``nilpotency'' accounts more accurately of property (\ref{period2}).}\end{footnote}
is totally equivalent to zero-regularization.

\section{Abelian $(2l+1)$-links invariants: a geometric computation}\label{4}

In this section we will show:

\vspace{5mm}

\noindent \textbf{Property 6}
\textit{In generalized CS theories, the only Wilson loops having non vanishing
expectation values are those of the homologically trivial links (modulo 2k). The expectation values of these Wilson loops are given by the self-linking of the corresponding link and the only required regularization is the one provided by framing (i.e. self-linking of the fundamental loops forming the link).}

\vspace{5mm}
We will first present the general ideas used to compute expectation values (\ref{WilsonGenMV}).
Then we will consider the particular case $M=S^{4l+3}$, the closest to the field theoretical
computation of section \ref{sect5}. We will next treat the less trivial case
$M=S^{2l+1} \times S^{2l+2}$.
In these two examples, we will present an alternative and more computational way to get
Property 6. Since $M$ is assumed without torsion, all its homology and
cohomology groups  are free and of finite type, {\it i.e} of the form $\mathbb{Z}^N$, for some
integer $N$. If $(\vec{e})_{I=1,...,N}$ denotes the canonical basis of $\mathbb{Z}^N$, then any
$\vec{u} \in \mathbb{Z}^N$ is written as
\[
\vec{u} = \sum_{I=1}^N u^I \vec{e}_I \, \, \, \, , \, \, \, \, u^I \in \mathbb{Z}.
\]

\subsection{Abelian $(2l+1)$-links invariants on $(4l+3)$-dimensional manifolds}

As already mentioned, $H_D^{2l+1} \left( {M,{\mathbb Z}} \right)^{\ast}$, as well as its smooth
version $H_D^{2l+1} \left( {M,{\mathbb Z}} \right)$, are affine bundles over the discrete space
$\check{H}^{2l+2}\left( {M,{\mathbb Z}} \right)$ . Although the Chern-Simons functional measure on
this space is written as in eqn. (\ref{20bis}), we need to give a more precise meaning to this
expression before we perform any computation. First, since the base space is of the form
$\mathbb{Z}^N$, the measure $d \mu_k(\omega)$ has to be decomposed into a sum of measures over each
(affine) fiber of $H_D^{2l+1} \left( {M,{\mathbb Z}} \right)^{\ast}$.
On each of these fibers we choose an origin, say $\omega_{\vec{u}}^0$, where
$\vec{u} \in \mathbb{Z}^N$ denotes the corresponding base point in
$\check{H}^{2l+2}\left( {M,{\mathbb Z}} \right)$.
Thus, $d \mu_k(\omega)$ reduces to a ``vectorial" measure on
$Hom\left( \Omega _{\mathbb Z}^{2l+2} \left( M \right),{{\mathbb R}/{\mathbb Z}} \right)$.
This amounts to pick up a global section for the affine bundle
$H_D^{2l+1} \left( {M,{\mathbb Z}} \right)^{\ast}$. The CS measure hence reads:
\begin{equation}
\label{21ter}
d \mu_k(\omega)
=
\sum_{\vec{u} \in \mathbb{Z}^N}  D \alpha
\exp \left\{ CS_k(\omega_{\vec{u}}^0 + \alpha) \right\}
=
\sum_{\vec{u} \in \mathbb{Z}^N} \, d \mu_k(\omega_{\vec{u}}^0; \alpha) \, ,
\end{equation}
where $\alpha \in Hom\left( \Omega _{\mathbb Z}^{2l+2}
\left( M \right),{{\mathbb R}/{\mathbb Z}} \right)$, $D \alpha$ is a measure on $Hom\left( \Omega _{\mathbb Z}^{2l+2} \left( M \right),{{\mathbb R}/{\mathbb Z}} \right)$, and each measure $d \mu_k(\omega_{\vec{u}}^0; \alpha)$ satisfies the Cameron-Martin property (\ref{20bis}).

On the other hand, inclusion (\ref{10bis}) together with Poincaré duality imply that on each
fiber of $H_D^{2l+1} \left( {M,{\mathbb Z}} \right)^{\ast}$ we can use, as an origin on this fiber,
a $(2l+1)$-cycle or equivalently its DB representative.
In particular, a fundamental
loop $\gamma_I^0$ can be associated with each basis vector $\vec{e}_I$ of $\mathbb{Z}^{N}$. Its DB representative $\eta_I^0$ then plays the role of origin on the fiber over $\vec{e}_I$. If $\vec{u} = \sum u^I \vec{e}_I$, then $\eta_{\vec{u}} \equiv \sum u^I \eta_I^0$ will be
a possible origin for the fiber over $\vec{u}$.
Note that the de Rham current of $\gamma_I^0$ would play the role of the ``curvature" of $\eta_I^0$,
as an element of
$Hom \left( \Omega^{2l+1} \left( M \right) / \Omega _{\mathbb Z}^{2l+1}
\left( M \right),{{\mathbb R}/{\mathbb Z}} \right)$.

Once such an origin for each fiber of $H_D^{2l+1} \left( {M,{\mathbb Z}} \right)^{\ast}$ has been
chosen, any DB class $\omega$ can be decomposed as
\begin{equation}
\label{DBdecomposition}
\omega
=
\sum_{I=1}^N u_{\omega}^I \eta_I^0 + \alpha
\equiv \vec{u}_{\omega} \cdot \vec{\eta}^{\, \, 0} + \alpha \, ,
\end{equation}
with $\alpha \in Hom\left( \Omega _{\mathbb Z}^{2l+2}
\left( M \right),{{\mathbb R}/{\mathbb Z}} \right)$, and $\vec{u}_{\omega}$
being the base point over which $\omega$ stands. In particular, the DB representative
$\eta$ of a cycle $\gamma$ will decompose as
\begin{equation}
\label{DBdecomposition2}
\eta =
\sum_{I=1}^N u_{\gamma}^I \eta_I^0 + \alpha
\equiv \vec{u}_{\gamma} \cdot \vec{\eta}^{\, \, 0} + \alpha \, .
\end{equation}

For a link $L$, we can express the expectation value of the corresponding Wilson line according
to our choice of basis $(\eta_I^0)_{I=1,...,N}$:
\bea
\label{21sept}
< W(L)>_{CS_{k}} = Z_k^{-1}
\sum\limits_{\vec{u}}  \int d \mu_k(\vec{u} \cdot \vec{\eta}^{\, \, 0}; \alpha) \,
 W(\vec{u},\alpha,\vec{v}_L,\beta) \, ,
\eea
where
\bea
\label{normalisateur}
Z_{k} =
\sum\limits_{\vec{u}} \int  d \mu_k(\vec{u} \cdot \vec{\eta}^{\, \, 0}; \alpha) \, ,
\eea
and
\bea
\label{Wilsondecomp}
W(\vec{u},\alpha,\vec{v}_L,\beta) =
\exp \left\{ 2i \pi \int_M (\vec{u} \cdot \vec{\eta}^{\, \, 0} +
\alpha) \ast_D (\vec{v}_{L} \cdot \vec{\eta}^{\, \, 0} + \beta) \right\} \,
\eea
is a rewriting of the Wilson line of $L$ with respect to the basis $(\eta_I^0)_{I=1,\ldots,N}$,  and
with the decomposition $\eta_L = \vec{v}_{L} \cdot \vec{\eta}^{\, \, 0} + \beta$ for  the DB
representative of $L$. We recall that $L$ is a link (a formal combination of charged fundamental
loops) hence a cycle.

Instead of evaluating the Wilson line (\ref{21sept}), we rather use the zero mode property. Let
$(\Sigma_{0}^{I})_{I=1,\ldots,N}$ be a collection of $(2l+2)$-cycles on $M$ which generates
$H_{2l+2}(M,\mathbb{Z})$ and are orthogonal to the fundamental loops $\gamma_I^0$:
\bea
\label{ortho}
\int_{\gamma_I^0} \beta_{0}^{J} = \delta_{IJ} = \Sigma_{0}^{J} \ti \gamma_I^0 \, ,
\eea
$\beta_{0}^{J}$ being the currents of the $\Sigma_{0}^{J}$, and \hspace{-2.5mm} $\ti$ \hspace{-2mm} denoting transversal intersection. Due to Poincaré and $Hom$ dualities there are as many $\beta_{0}^{J}$ as $\gamma_I^0$.

Let us consider again:
\bea
\label{newmeanvalue}
< W(L)>_{CS_{k}} = Z_k^{-1}
\int d \mu_k(\omega) \exp \left\{ 2i\pi \int_{L} \omega \right\} \, ,
\eea
into which we perform the shift
\bea
\label{shiftzeromode}
\omega \rightarrow \omega + \sum_{I=1}^{N} m_I \frac{\beta_{0}^{I}}{2k} \, ,
\eea
for a collection of integers $m_I$. This gives:
\bea
\label{meanvalueshifted}
< W(L)>_{CS_{k}}
= Z_k^{-1}
\int d \mu_k( \omega + \sum_{I=1}^{N} m_I \frac{\beta_{0}^{I}}{2k})
\exp
 \left\{ 2i\pi \int_{L} \left(\omega + \sum_{I=1}^{N} m_I
\frac{\beta_{0}^{I}}{2k} \right)
\right\} \, .
\eea
Using Property 3, we obtain:
\bea
\label{meanvalueshifted}
< W(L)>_{CS_{k}}
= Z_k^{-1}
\int d \mu_k( \omega ) \exp \left\{  2i\pi \int_{L} \omega \right\}
\exp \left\{ 2i \pi \sum_{I=1}^{N} \frac{m_I}{2k} \int_{L} \beta_{0}^{I}  \right\}  \, .
\eea
That is to say:
\bea
\label{meanvalueshifted2}
< W(L)>_{CS_{k}} = < W(L)>_{CS_{k}} \exp \left\{ 2i \pi \sum_{I=1}^{N} \frac{m_I}{2k} \int_{L} \beta_{0}^{I}  \right\}  \, .
\eea
Since this has to hold for any collection of integers $(m_I)_{I=1,\ldots,N}$, we conclude that,
for a non vanishing mean value:
\bea
\label{constraint}
\int_{L} \beta_{0}^{I} = 0 \, \, [2k],
\eea
$\forall I \in \left\{1,\ldots,N \right\}$.
Thus, if we forget about $[2k]$, the link $L$ has to be "orthogonal" to the generators of $H_{2l+2}(M,\mathbb{Z})$, which means that $L$
must be homologically trivial, for the mean value of the corresponding Wilson loop to be non vanishing. When $L$ is not trivial, the mean
value of the Wilson loop it defines has to be zero. The modulo $2k$ appearing in eqn. (\ref{constraint}) simply reminds us of the $2k$-nilpotency
property (\ref{period2}).

Finally, let $L$ be an homologically trivial link in $M$. This amounts to set $\vec{v}_L = \vec{0}$ in eqn. (\ref{21sept}), thus reducing it to:
\bea
\label{simplest}
\sum\limits_{\vec{u}}  \int D \alpha
\exp \left\{ CS_k(\vec{u} \cdot \vec{\eta}^{\, \, 0} +
\alpha) \right\} \exp \left\{ 2i \pi \int_M (\vec{u} \cdot \vec{\eta}^{\, \, 0} + \alpha) \ast_D \beta_L \right\} \, ,
\eea
where $\beta_L$ is the DB class of a current of a $(2l+2)$-chain with boundary $L$. Now let us perform into eqn. (\ref{simplest}) the shift:
\bea
\label{shift1}
\alpha \rightarrow \alpha + \frac{\beta_L}{2k} \, ,
\eea
what leads to:
\bea
\label{simplestshifted}
\sum\limits_{\vec{u}}  \int D \alpha
\exp \left\{ CS_k(\vec{u} \cdot \vec{\eta}^{\, \, 0} +
\alpha) \right\}
\exp \left\{ - 2i \pi k \int_M \frac{\beta_L}{2k} \ast_D \frac{\beta_L}{2k} \right\} \, .
\eea
Hence, we obtain:
\bea
\label{finally}
< W(L)>_{CS_{k}} = \exp \left\{ - \frac{2i \pi}{4k}  \int_M \beta_L \wedge d \beta_L \right\} \, .
\eea
The integral in this expression is, modulo zero-regularization via framing, exactly the self-linking number of the link $L$ \cite{SLF,Po,JHW}, itself made of self-linking (defined via framing) and linking of the fundamental loops composing $L$. We stress out that while the link has to be homologically trivial, its components do not have to. This completes the proof of Property 6.

Of course we could have directly used property (\ref{21quart}) together with the shift (\ref{shift1}) to obtain
eqn. (\ref{finally}). However we have preferred to use the explicit definition (\ref{21ter}) of the functional
integral rather than the formal one.

Let us have a closer look at a first example where zero modes are not required to be treated: the spheres. This will provide us with a general property concerning $(4l+3)$-manifolds whose $(2l+1)$-th homology group
vanishes.

\subsection{Abelian links invariants on $S^{4l+3}$}

Since $\check{H}^{2l+2}\left( {S^{4l+3},{\mathbb Z}}
\right) = 0 = \check{H}^{2l+1}\left( {S^{4l+3},{\mathbb Z}}
\right)$, the first of the exact sequences (\ref{1bis}) reduces to:
\bea
\label{22bis}
H_D^{2l+1} \left( {S^{4l+3},{\mathbb Z}} \right)
&\simeq&
{\Omega^{2l+1}\left( S^{4l+3} \right)} \mathord{\left/
{\vphantom {{\Omega^p\left( S^{4l+3} \right)} {\Omega_{\mathbb Z}^{2l+1} \left( S^{4l+3}
\right)}}} \right. \kern-\nulldelimiterspace} {\Omega_{\mathbb Z}^{2l+1} \left( S^{4l+3}
\right)} \\
&=&
\Omega^{2l+1}\left( S^{4l+3} \right) / d \Omega^{2l}\left( S^{4l+3} \right) \, , \nonumber
\eea
and the dual sequence (\ref{8bis}) to:
\bea
\label{23bis}
H_D^{2l+1} \left( {S^{4l+3},{\mathbb Z}}\right)^{\ast}
&\simeq&
Hom\left( \Omega _{\mathbb Z}^{2l+2} \left( S^{4l+3} \right),{{\mathbb R}/{\mathbb Z}} \right) \\
&=&
Hom\left( d \Omega^{2l+1} \left( S^{4l+3} \right),{{\mathbb R}/{\mathbb Z}} \right) \, . \nonumber
\eea
These isomorphisms are somehow canonical if we consider that the choice of the zero class,
$\textbf{0}$, as origin of these spaces is canonical. More explicitly, for any
$\omega \in H_D^{2l+1} \left( {S^{4l+3},{\mathbb Z}} \right)^{\ast}$ there is a
$\alpha \in Hom\left( \Omega _{\mathbb Z}^{2l+2}
\left( S^{4l+3} \right),{{\mathbb R}/{\mathbb Z}} \right)$ such that:

\begin{equation}
\label{DBLink}
\omega = \textbf{0} + \alpha \equiv \alpha \, ,
\end{equation}

\noindent This corresponds to choose the zero cycle $z \equiv 0$ as origin, the DB representative
of this cycle being $\textbf{0}$. Since $\check{H}_{2l+1}\left( {S^{4l+3},{\mathbb Z}}\right) =
0$, any $(2l+1)$-cycle in $S^{4l+3}$ is trivial, \textit{i.e.} a boundary. Hence, if $L$
denotes a $(2l+1)$-link which is the sum of charged fundamental $(2l+1)$-loops $\gamma_i^0$ on
$S^{4l+3}$:
\begin{equation}
\label{24bis}
L = \sum_{i=1}^N q_i \gamma_i^0 \, ,
\end{equation}
then there exists some $(2l+2)$-chain, $\Sigma_L$, such that $L = b \Sigma_L$. Geometrically,
$\Sigma_L$ can be seen as a $(2l+2)$-surface in $S^{4l+3}$. This surface is of course not unique,
but two of them only differ by a closed $(2l+2)$-surface. As explained in \cite{BGST}, the de Rham
current of such a $\Sigma_L$, $\beta_\Sigma$, completely determines the DB representative,
$\eta_L$, of $L$, according to:
\begin{equation}
\label{DBLink}
\eta_L = \textbf{0} + \beta_\Sigma \, ,
\end{equation}
with $\beta_\Sigma \in
Hom\left( \Omega _{\mathbb Z}^{2l+2} \left( S^{4l+3} \right),{{\mathbb R}/{\mathbb Z}} \right)$.
The Wilson line of $L$ is then written:
\begin{equation}
\label{WilsonTriv}
W(\alpha,L) =  \exp \left\{ 2i \pi \int_{S^{4l+3}} \alpha \ast_D \beta_\Sigma \right\} \, ,
\end{equation}
and its expectation value reads:
\begin{equation}
\label{25bis}
< W(L)>_{CS_k} =   \frac{ \int D \alpha  \exp \left\{ 2i \pi
k \int_{S^{4l+3}} \alpha \ast_D \alpha + 2i \pi \int_{S^{4l+3}} \alpha \ast_D \beta_\Sigma \right\}  }{\int D \alpha  \exp \left\{ 2i \pi
k \int_{S^{4l+3}} \alpha \ast_D \alpha  \right\} } \, .
\end{equation}
Seen as an element of $Hom\left( \Omega _{\mathbb Z}^{2l+2} \left( S^{4l+3} \right),{{\mathbb R}/{\mathbb Z}} \right)$, $\beta_\Sigma / 2k$ fulfills:
\begin{equation}
\label{2kcurrent}
2k (\frac{\beta_\Sigma}{2k}) = \beta_\Sigma \, .
\end{equation}
However, the corresponding DB class, $\textbf{0} + (\beta_\Sigma / 2k)$, is not the representative of any fundamental loop in $S^{4l+3}$.

Next, we perform the change of variable:
\begin{equation}
\label{trivshift}
\alpha \mapsto \widetilde{\alpha} = \alpha + \frac{\beta_\Sigma}{2k} \, ,
\end{equation}
into eqn. (\ref{25bis}). This turns the expectation value into:
\begin{equation}
\label{Wilsonfinal}
< W(L)>_{CS_k} =     \exp \left\{ - 2i \pi
k \int_{S^{4l+3}} \frac{\beta_\Sigma}{2k} \ast_D \frac{\beta_\Sigma}{2k} \right\}  \, .
\end{equation}
Making explicit the DB product within this expression, we obtain:
\begin{equation}
\label{Wilsonfinal}
< W(L)>_{CS_k} = \exp \left\{ - \frac{2i \pi}{4k}
 \int_{S^{4l+3}} \beta_\Sigma \wedge d \beta_\Sigma \right\}  \, ,
\end{equation}
what is exactly eqn. (\ref{finally}).

Finally in terms of the charged fundamental loops, $\gamma_i^0$, building $L$, we have
\begin{equation}
\label{Wilsonfinal2}
< W(L)>_{CS_k} =  \exp \left\{ - \frac{2i \pi}{4k}
\sum_{i,j} q_i L(\gamma_i^0,\gamma_j^0) q_j \right\}  \, ,
\end{equation}
where $L(\gamma_i^0,\gamma_j^0)$ is the linking number of $\gamma_i^0$ with $\gamma_j^0$, that is
to say:
\begin{equation}
\label{GenLinking}
L(\gamma_i^0,\gamma_j^0) = \int_{S^{4l+3}} \alpha_i^0 \wedge d \alpha_j^0 \, ,
\end{equation}
with $\alpha_i^0$ the de Rham current for which $\textbf{0} + \alpha_i^0$ is the DB
representative of the fundamental loop $\gamma_i^0$. As for ``diagonal" terms
$L(\gamma_i^0,\gamma_i^0)$ we regularize them using the usual framing procedure (what we have called
zero-regularization):
\begin{equation}
\label{GenLinking}
L(\gamma_i^0,\gamma_i^0) \equiv L(\gamma_i^0,\gamma_i^{0f}) \, .
\end{equation}
As in the three dimensional case extensively detailed in \cite{GT}, the abelian invariants thus obtained are
nothing but those coming from linking and self-linking numbers, that is to say intersection theory in
$S^{4l+3}$. Let's note that this result is what we are supposed to recover via a quantum field theory
approach. There, the gauge fixing procedure is supposed to provide a choice of representatives for DB
classes, and the propagator thus obtained appears like an inverse of the de Rham differential $d$, deeply
related to the Poincaré chain homotopy operator. The consistency of the procedure is ensured by the fact
that if $\gamma$ is a loop (a $(2l+1)$-cycle), and if $\Sigma$ is a $(2l+2)$-chain such that $b \Sigma =
\gamma$, which corresponds to $d \beta_{\Sigma} = \eta_{\gamma}$ in term of currents, then $\beta_{\Sigma}$
(as the current of an integral chain) is unique up to closed $(2l+1)$-currents (of integral
$(2l+2)$-cycles). However, on $S^{4l+3}$ any $(2l+2)$-cycle is trivial so $\beta_{\Sigma}$ is unique up to
$d \chi$, where $\chi$ is the $2l$-current of an arbitrary $(2l)$-chain. This means $d \beta_{\Sigma} =
\eta_{\gamma}$ has to be inverted on classes $\beta_{\Sigma} \sim \beta_{\Sigma} + d \chi$. This is exactly
gauge invariance from the point of view of integral chains (and currents). This will be detailed in section
\ref{sect5}.

What we have done here for $S^{4l+3}$ can be straightforwardly applied to any $(4l+3)$-manifold
$M$ for which
$\check{H}^{2l+1}\left( {M,{\mathbb Z}}\right) = 0 =
\check{H}^{2l+2}\left( {M,{\mathbb Z}} \right)$, leading to exactly the same final result.

\vspace{5mm}

\noindent \textbf{Property 7} \textit{Over a $(4l+3)$-dimensional closed manifold, without
torsion, whose $(2l+1)$th homology groups vanishes, the generalized abelian Wilson loop of a link $L$
defines a link invariant made of the self-linkings, the  linkings and the charges of the fundamental loops composing $L$.}

\vspace{5mm}

The second example will present a homologically non trivial case which is the equivalent of the three
dimensional pedagogical case $S^1 \times S^2$ widely discussed in \cite{GT}.

\subsection{Abelian links invariants on $S^{2l+1} \times S^{2l+2}$}

Let us now consider the less trivial case $M \equiv S^{2l+1} \times S^{2l+2}$ for which
$\check{H}^{2l+2}\left( M,{\mathbb Z} \right) = \mathbb{Z} =
\check{H}^{2l+1}\left( M,{\mathbb Z}\right)$, so that:
\bea
\label{notrivDB}
H_D^{2l+1} \left( {M,{\mathbb Z}}
\right) &\simeq& \mathbb{Z} \times
\frac{\Omega^{2l+1}\left( M \right)}{\Omega_{\mathbb Z}^{2l+1} \left( M
\right)} \, ,
\eea
and:
\bea
\label{notrivDBstar}
H_D^{2l+1} \left( {M,{\mathbb Z}}\right)^{\ast} &\simeq& \mathbb{Z}
\times
Hom\left( \Omega _{\mathbb Z}^{2l+2} \left( M \right),{{\mathbb R}/{\mathbb Z}} \right) \, ,
\eea
none of these isomorphisms being canonical. However, over the base point $0 \in \mathbb{Z}$ we still have the zero DB class (which is again the representative of the zero cycle in $M$), so that this particular fiber of $H_D^{2l+1} \left( {M,{\mathbb Z}}\right)^{\ast}$ can be (almost canonically) identified with the translation group
$Hom\left( \Omega _{\mathbb Z}^{2l+2} \left( M \right),{{\mathbb R}/{\mathbb Z}} \right)$. This is similar to what previously happened in the case of the sphere $S^{(4l+3)}$.
However, we now have $\check{H}_{2l+1}\left( M,{\mathbb Z} \right) = \mathbb{Z}$, which means that there are non trivial $(2l+1)$-loops in $M$. Accordingly, we pick up a fundamental $(2l+1)$-loop $\gamma^0$ which generates $\check{H}_{2l+1}\left( M,{\mathbb Z} \right)$. Formally $\gamma^0$ is given by a $S^{2l+1}$ in $M$. Its DB representative, $\eta^0$ will play the role of the origin on the fiber over $1 \in \mathbb{Z}$ in $H_D^{2l+1} \left( {M,{\mathbb Z}}\right)^{\ast}$. If $L$ is a link in $M$, then its $DB$ representative, $\eta_L$, satisfies
\bea
\eta_L = n_L \eta^0 + \beta_\Sigma \, ,
\eea
with $n_L \in \mathbb{Z}$ the base point over which $\eta_L$ stands in
$H_D^{2l+1} \left( {M,{\mathbb Z}}\right)^{\ast}$, and the translation term
$\beta_\Sigma$ belongs to $Hom\left( \Omega _{\mathbb Z}^{2l+2}
\left( M \right),{{\mathbb R}/{\mathbb Z}} \right)$. Once more, $\beta_\Sigma$ alternatively denotes the de Rham current of a $(2l+2)$-chain $\Sigma_L$ for which $L = n_L \gamma^0 + b \Sigma_L$ as well as the DB class this current defines via sequence (\ref{8bis}). Such a
chain is not unique, but two of them differ by a  $(2l+2)$-cycle whose de Rham current belongs to
the zero class in
$Hom\left( \Omega _{\mathbb Z}^{2l+2} \left( M \right),{{\mathbb R}/{\mathbb Z}} \right)$,
making $\beta_\Sigma$ unique from the DB class point of view.

So, up to the normalization factor $Z^{-1}_k$, the expectation value
(\ref{21sept}) reduces to:
\bea
\label{notrivWilson}
\sum_{m \in \mathbb{Z}} \int D \alpha
\exp \left\{ 2i \pi \int_M  (m \eta^0 + \alpha) \ast_D (km \eta^0 +
k \alpha + n_L \eta^0 + \beta_\Sigma) \right\}  \, .
\eea
Instead of using the elegant zero-mode property, as was done to establish Property 6, we shall present a somehow more computational approach. Although this will be a bit "heavier", we make this choice in order to show more explicitly the usefulness of zero modes as well as of zero-regularization.

Since it provides the final answer, let us first consider the case where $n_L = 0$ ( \textit{i.e.} when $L$ is homologically trivial).
Then expression (\ref{notrivWilson}) takes the form:
\bea
\label{notrivWilsontriv}
\sum_{m \in \mathbb{Z}} \int D \alpha
\exp \left\{ 2i \pi \int_M  (m \eta^0 +
\alpha) \ast_D (km \eta^0 + k \alpha + \beta_\Sigma) \right\}  \, .
\eea
For the same reasons than in the previous example,
$\beta_\Sigma / 2k \in
Hom\left( \Omega _{\mathbb Z}^{2l+2} \left( M \right),{{\mathbb R}/{\mathbb Z}} \right)$.
So, we perform the shift:
\begin{equation}
\label{notrivshift}
\alpha \mapsto \widetilde{\alpha} = \alpha + \frac{\beta_\Sigma}{2k} \, .
\end{equation}
The expectation value of the Wilson line of $L$ then simplifies into:
\bea
\label{notrivWilsontriv2}
\sum_{m \in \mathbb{Z}} \int D \alpha
\exp \left\{ 2i \pi \int_M  (m \eta^0 + \alpha) \ast_D (km \eta^0 +
k \alpha) \right\} \hspace{5cm} \\
\times \exp \left\{ - 2i \pi \int_M \frac{\beta_\Sigma}{2k} \ast_D \frac{\beta_\Sigma}{2k} \right\} \, ,
\nonumber
\eea
that is to say:
\bea
\label{expectWilson}
< W(L) >_{CS_k}= \exp \left\{ - 2i \pi k
\int_M \frac{\beta_\Sigma}{2k} \ast_D \frac{\beta_\Sigma}{2k} \right\} \, ,
\eea
or equivalently:
\bea
\label{expectWilsonfinal}
< W(L) >_{CS_k}= \exp \left\{ - \frac{2i \pi}{4k}
\int_M \beta_\Sigma \wedge d \beta_\Sigma \right\} \, ,
\eea
just as in the $S^{4l+3}$ case. Once more, this is totally similar to what happens in the three dimensional case $S^1 \times S^2$ detailed
 in \cite{GT}. This turns out to be the same expression as eqn. (\ref{Wilsonfinal}), and of course as eqn. (\ref{finally}): the link invariant is
made of linking and self-linking numbers of the fundamental loops forming the link. However let us stress again that whereas the link $L$ has to be homologically trivial, this is not the case of its components.

Let us now assume that $n_L$ is not zero (nor an integral multiple of $2k$, although this can be dealt with straightforwardly). If we expand all the
expressions within the exponentials appearing in eqn. (\ref{notrivWilson}), and then apply the
zero-regularization to $\eta^0  \ast_D \eta^0$, we obtain the expression:
\bea
\label{notrivWilson2}
 k \alpha \ast_D \alpha + \alpha \ast_D \beta_\Sigma +
(2km + n_L) \eta^0 \ast_D \alpha + m \eta^0 \ast_D \beta_\Sigma \, .
\eea
Once more, we perform the shift (\ref{notrivshift}), and get, after some simplifications:
\bea
\label{notrivWilsonshifted}
 k \alpha \ast_D \alpha +
(2km + n_L) \eta^0 \ast_D \alpha - k \frac{\beta_\Sigma}{2k}
\ast_D \frac{\beta_\Sigma}{2k} - n_L \eta^0 \ast_D \frac{\beta_\Sigma}{2k} \, .
\eea
The last two terms are independent of $m$ and $\alpha$, and then give rise to:
\bea
\label{facorized}
\exp \left\{ - 2i \pi  \int_M \frac{\beta_\Sigma}{2k}
\ast_D ( k \frac{\beta_\Sigma}{2k} + n_L \eta^0 ) \right\} \, ,
\eea
out of the integration and sum in eqn. (\ref{notrivWilsontriv}). In the remaining factor, we can
invert the sum over $m$ with the integration over $\alpha$, thus obtaining:
\bea
\label{notrivWilsonSum}
\int D \alpha e^{2i \pi k \int_M \alpha \ast_D \alpha}
\sum_{m \in \mathbb{Z}} \exp \left\{ 2i \pi \int_M
\left(  (2km + n_L)  \eta^0 \ast_D \alpha \right) \right\}  \, .
\eea
But:
\bea
\label{algebre}
\sum_{m \in \mathbb{Z}}
\exp \left\{ 2i \pi \int_M
\left(  (2k)m  \eta^0 \ast_D \alpha \right) \right\}
&=&
\sum_{m \in \mathbb{Z}} \exp \left\{ 2i \pi (2km) \int_{\gamma^0} \alpha  \right\} \\
&=&
\sum_{K \in \mathbb{Z}} \delta \left( \int_{\gamma^0} \alpha -  K/2k \right)  \nonumber \, .
\eea

\noindent
Putting this back into eqn. (\ref{notrivWilsonSum}), and performing some algebraic juggling, we obtain:
\bea
\label{notrivWilson3}
\sum_{K \in \mathbb{Z}} e^{ 2i \pi n_L K/2k}
 \int D \alpha \, \, \delta
\left( \int_{\gamma^0} \alpha -  K/2k \right) e^{2i \pi k \int_M \alpha \ast_D \alpha} \,.
\eea
Let us introduce a closed $(2l+2)$-surface $\Sigma_0$, with de Rham $(2l+1)$-current $\rho^0$, which satisfies:
\bea
\label{zeromodes}
\int_{\gamma^0} \rho^0 = 1 = \Sigma_0 \ti \gamma^0 \, .
\eea
This surface is a generator of $\check{H}^{2l+1}\left( M,{\mathbb Z}\right) \simeq \check{H}_{2l+2}\left( M,{\mathbb Z}\right) = \mathbb{Z}$ and is formally a sphere $S^{(2l+2)}$ in $M = S^{(2l+1)} \times S^{(2l+2)}$. The
(trivial) DB class associated with $\rho^0$ ( also denoted $\rho^0$) give rises to the DB class $\rho^0 / 2k$, which is non trivial since:
\bea
\label{zeromodes2}
\int_{\gamma^0} \frac{\rho^0}{2k} \, \mathop = \limits_\mathbb{Z} \, \frac{1}{2k} \, .
\eea
Actually, $\rho^0 / 2k \in
Hom\left( \Omega _{\mathbb Z}^{2l+2} \left( M \right),{{\mathbb R}/{\mathbb Z}} \right)$
and the DB class it determines is $\textbf{0} + \rho^0 / 2k$. Moreover, as seen when establishing the zero-mode property:
\bea
\label{rhoproperty}
\int_M \frac{\rho^0}{2k} \ast_D \frac{\rho^0}{2k} \, \, \mathop = \limits_\mathbb{Z} \, \, 0 \, \, \mathop = \limits_\mathbb{Z} \, \, 2k \int_M \frac{\rho^0}{2k} \ast_D \alpha \, ,
\eea
for any $\alpha \in
Hom\left( \Omega _{\mathbb Z}^{2l+2} \left( M \right),{{\mathbb R}/{\mathbb Z}} \right)$. Consequently, eqn. (\ref{notrivWilson3}) reads:
\bea
\label{notrivWilson4}
\sum_{K \in \mathbb{Z}} e^{ 2i \pi n_L K/2k}
 \int D \alpha \, \, \delta \left( \int_{\gamma^0}
(\alpha -  K \frac{\rho^0}{2k}) \right) e^{2i \pi k \int_M \alpha \ast_D \alpha} \,,
\eea
and for each value of $K$, if we perform the shift:
\begin{equation}
\label{notrivshift2}
\alpha \mapsto \widetilde{\alpha} = \alpha - K \frac{\rho^0}{2k} \, ,
\end{equation}
and use eqn. (\ref{rhoproperty}), the expression under the integral in eqn. (\ref{notrivWilson3}) turns out to be independent of $K$. Thus:
\bea
\label{Sumout}
\sum_{K \in \mathbb{Z}} e^{ 2i \pi n_L K/2k} \, ,
\eea
factorizes out of eqn. (\ref{notrivWilson4}).
The same procedure has to be applied to the denominator of expression (\ref{21sept}) (which is the
normalization factor needed to compute expectation values), producing a term:
\bea
\label{Sumout2}
\sum_{K \in \mathbb{Z}} 1 \, .
\eea
None of the expressions (\ref{Sumout}) and (\ref{Sumout2}) is well-defined. However, using $2k$-nilpotency, we can reduce each of these infinite sums to a sum over a period, thus obtaining:
\bea
\label{Sumout3}
\sum_{K = 0}^{2k-1} e^{ 2i \pi n_L K/2k} =
\left| \begin{array}{rr}
 2k &   \textrm{if} \, \, \, n_L = 0 \\
 0 & \textrm{otherwise}
\end{array} \right.
\eea
for the former one and
\bea
\label{Sumout4}
\sum_{K = 0}^{2k-1} 1 = 2k \, .
\eea
for the latter one. The ``regularized" quotient defining the expectation value will then be taken
as:
\bea
\label{regulsum}
\mathop {\textrm{lim}} \limits_{N \mapsto \infty}
\frac{N \sum_{K = 0}^{2k-1} e^{ 2i \pi n_L K/2k}}{N \sum_{K = 0}^{2k-1} 1}
=
\frac{\sum_{K = 0}^{2k-1} e^{ 2i \pi n_L K/2k}}{\sum_{K = 0}^{2k-1} 1} =
\left| \begin{array}{rr}
 1 &   \textrm{if} \, \, \, n_L = 0 \, \, [2k] \\
 0 & \textrm{otherwise} \, . \, \, \, \, \, \, \, \, \, \, \,
\end{array} \right.
\eea

Hence, when $n_L \neq 0 \, \, [2k]$, the expectation value of the corresponding Wilson line is
zero, while when $n_L = 0$ the expectation value is given by eqn. (\ref{expectWilson}). Due to
$2k$-nilpotency, when $n_L = 2kN$, with $N \in \mathbb{Z}^\ast$, then the corresponding link
invariant is trivial. These results are a clear generalization of those investigated in \cite{GT}
for the three dimensional case. Also, it is quite obvious how to deal with a more general case than the quite simple product $S^{2l+1} \times S^{2l+2}$, as long as $M$ is torsionless. The case of $(4l+3)$-manifolds with torsion might be treated extending \cite{Th}.

\section{Naive abelian gauge field theory and $(2l+1)$-links invariants}\label{sect5}

This section provides a formulation of the abelian $(4l+3)$-dimensional Chern Simons theory on
$\mathbb{R}^{4l+3}$ with Euclidean metric in terms of a lagrangian density involving a $U(1)$ connection
\textit{i.e.} gauge field $A$, plus gauge fixing. This formulation, coined ``naive gauge field theory" extends
eqns. (\ref{11bis}), (\ref{12bis}) to the $(4l+3)$-dimensional case, and is the one familiar to field
theorists. The presentation is formulated in a somewhat hybrid way conveniently using
notations which keep track of the geometric nature of the fields and operations, combined with
algebraic manipulations familiar in field theory. We aim here at emphasizing the ambiguities or
weaknesses arising in this framework, in order to stress where the above non perturbative formulation in
terms of DB cohomology classes brings clarification.
In particular, the normalization of both the level $k$ and loop charges $e$ are a priori
unspecified in the naive field theory approach: the prescription that they have to be integers is
{\em ad hoc}, whereas they are bound to be integers \textit{ab-initio} in the DB approach.
Furthermore, the naive approach leads to ill-defined self-linking integrals which require to be given
meaning and integer values by some extrinsic regularization
procedure, such as framing, whereas the DB approach was shown above provides a natural
\textit{regularization independent normalization} prescription for the latter.
Last, this study on $\mathbb{R}^{4l+3}$ also suggests which complications may arise when trying to
extend the naive field theoretical framework to manifolds with non trivial cohomology.

\subsection{Formulation and computation on $\mathbb{R}^{4l+3}$}

\noindent
The lagrangian density\footnote{Properly speaking the Chern-Simons lagrangian {\em density} familiar
to field theorists is the Hodge $^{*}$ dual (on $\mathbb{R}^{4l+3}$ with Euclidean metric) of
the lagrangian {\em $(4l+3)$-form} familiar to geometers introduced by eq. (\ref{11bis}).
The left hand side of eq. (\ref{lagr-51}) should thus be
$\mbox{}^{*}{\cal L}_{CS} \left( A^{(2l+1)} \right)$, and likewise for the gauge fixing
lagrangian density ${\cal L}_{GF}$ in the forthcoming subsection \ref{subsect511}.
This sloppiness will hopefully not be confusing.} ${\cal L}_{CS} \left( A^{(2l+1)} \right)$ of the
abelian $(4l+3)$-dimensional Chern-Simons theory reads:
\begin{equation}\label{lagr-51}
{\cal L}_{CS} \left( A^{(2l+1)} \right) = \frac{1}{2} A^{(2l+1)} \wedge d \, A^{(2l+1)} \, .
\end{equation}
An extra factor 1/2 is introduced in the normalization of ${\cal L}_{CS}$ with respect to the
normalization of $cs_1(A)$ in eq. (\ref{11bis}).
This normalization choice is convenient to calculate the propagator
of the $A^{(2l+1)}$ field. This extra factor is subsequently compensated by defining the
Chern Simons action as $4 i \pi$ times the integral of ${\cal L}_{CS}$ indeed matching
the normalization of $CS_1(A)$ in eq. (\ref{12bis}).

\vspace{0.3cm}

\noindent
The degeneracy coming from the gauge invariance $A^{(2l+1)} \to A^{(2l+1)} + d\, \Lambda^{(2l)}$ of
this lagrangian density shall be fixed, in order that the functional integral giving the
generating functional, and, in particular, the propagator of the $A^{(2l+1)}$ field be defined.

\subsubsection{Covariant gauge fixing and corresponding propagator}\label{subsect511}

In the three dimensional case, a common procedure consists in imposing the ``covariant gauge fixing"
$d\,^{*}A^{(3)} = 0$ by adding the following Lagrange constraint:
\begin{equation}\label{gauge-fix-3d}
{\cal L}_{GF}^{(3d)} = B^{(0)} \wedge d\,^{*}A^{(3)}
\end{equation}
where $^{*}$ here denotes the Hodge dual operation with respect to the Euclidean metric
on $\mathbb{R}^{3}$ and the Lagrange multiplier $B^{(0)}$ is a scalar field {\em i.e.} a zero-form.
Let from now on $^{*}$ denote the Hodge dual operation on flat Euclidean $\mathbb{R}^{4l+3}$,
such that for any $q$-form $B^{(q)}$,
$^{**}B^{(q)} = (-1)^{q(4l+3-q)}B^{(q)} = B^{(q)}$.
The naive straightforward generalization
of eqn. (\ref{gauge-fix-3d}) by means of a single auxiliary $2l$-form $B^{(2l)}$ according to
\[
{\cal L}_{GF}^{naive} =  B^{(2l)} \wedge d\,^{*}A^{(2l+1)}
\]
is not effective as ${\cal L}_{GF}^{naive}$ still has the residual gauge invariance
$B^{(2l)} \to B^{(2l)} + d\, \Lambda^{(2l-1)}$. An appropriate formulation
requires a collection of $2l+1$ auxiliary forms of decreasing degrees
$(B^{(2l)}, B^{(2l-1)}, \cdots, B^{(0)})$, according to:
\begin{equation}
{\cal L}_{GF} =
B^{(2l)} \wedge d\,^{*}A^{(2l+1)} \, + \, B^{(2l-1)} \wedge d\,^{*}B^{(2l)}
\, + \, \cdots\, \, + \, B^{(0)} \wedge d\,^{*}B^{(1)}\, .
\end{equation}
Regrouping all the fields into \[
\vec{\cal A}  =({\cal A}_{1},{\cal A}_{2}, {\cal A}_{3},\cdots,{\cal A}_{2l+2}) \equiv
(A^{(2l+1)}, B^{(2l)}, B^{(2l-1)}, \cdots, B^{(0)})
\]
we can compactly write the full action given by
${\cal L}_{tot}= {\cal L}_{CS} \left( A^{(2l+1)} \right) + {\cal L}_{GF}$ as a scalar product:
\begin{equation}
\int {\cal L}_{tot} = \int_{\mathbb{R}^{4l+3}} \vec{\cal A} \wedge \, ^{*} D \vec{\cal A}
\equiv
\frac{1}{2} \left( \vec{\cal A} , D  \vec{\cal A} \right)
\end{equation}
with:
\begin{equation}
D =
\left[
\begin{array}{rrrrrrrrrrr}
 ^{*}d  &   -d   &   0    &  0 &     &        &        &    &     &        &  \\
 \delta &    0   &   d    &  0 &     &        &        &    &     &        &  \\
    0   & \delta &   0    & -d &     &        &        &    &     &        &  \\
    0   &    0   & \delta &  0 &     &        &        &    &     &        &  \\
        &        &        &    & ... &        &        &    &     &        &  \\
        &        &        &    &     &   0    &   d    &  0 &     &        &  \\
        &        &        &    &     & \delta &   0    & -d &     &        &  \\
        &        &        &    &     &   0    & \delta &  0 &     &        &  \\
        &        &        &    &     &        &        &    & ... &        &  \\
        &        &        &    &     &        &        &    &     &   0    & \ - d  \\
        &        &        &    &     &        &        &    &     & \delta &     0
\end{array}
\right]
\end{equation}
where $\delta \, \equiv \! ^{*}d^{\,*}$ is the co-differential associated with the Hodge dual.
The Euler-Lagrange equations of motion of the $\vec{\cal A}$ field read:
\begin{equation}
D \vec{\cal A} = 0
\end{equation}
The propagator $<\vec{\cal A}(x) \otimes \vec{\cal A}(y)>$ of the field $\vec{\cal A}$ is
the inverse of the operator $D$ conveniently determined solving
\begin{equation}\label{eq-propag-def}
D \,  <\vec{\cal A}(x) \otimes \vec{\cal A}(y)> =
\delta^{(4l+3)}(x-y) \, 1\!\mbox{I}_{2l+2}
\end{equation}
by means of Fourier transformation, taking advantage of translation invariance on Euclidean space
$\mathbb{R}^{4l+3}$. It is especially convenient to use a Fourier transformation, defined by means
of Berezin integration, which preserves the degrees of forms, as detailed in Appendix A.
The Fourier transform of $D \, \delta^{(4l+3)}(x-y)$ reads:
\begin{equation}\label{fourier-D}
\overrightharpoon{D}
\, = \,
-i \,\left[
 \begin{array}{rrrrrrrrrrr}
 ^{*}P  &   -P   &   0    &  0 &     &        &        &    &     &        &  \\
  \Xi   &    0   &   P    &  0 &     &        &        &    &     &        &  \\
    0   &  \Xi   &   0    & -P &     &        &        &    &     &        &  \\
    0   &    0   & \Xi    &  0 &     &        &        &    &     &        &  \\
        &        &        &    & ... &        &        &    &     &        &  \\
        &        &        &    &     &   0    &   P    &  0 &     &        &  \\
        &        &        &    &     & \Xi    &   0    & -P &     &        &  \\
        &        &        &    &     &   0    & \Xi    &  0 &     &        &  \\
        &        &        &    &     &        &        &    & ... &        &  \\
        &        &        &    &     &        &        &    &     &   0    & - P  \\
        &        &        &    &     &        &        &    &     & \Xi    &   0
 \end{array}
\right]\, .
\end{equation}
The expression for $P$ and $\Xi$ are given in eqns. (\ref{PetXi}) of Appendix A.

The Fourier transforms $\overrightharpoon{N}_{jk}$ of the
$<{\cal A}_{2l+2-j} \otimes {\cal A}_{2l+2-k}>$ satisfy:
\begin{eqnarray}
-i \, \left(
 ^{*}P \, \wedge \overrightharpoon{N}_{1,j}-P \, \wedge \overrightharpoon{N}_{2,j}
\right)
& = &
\delta_{1,j} \, \mathrm{Id}_{(2l+1)} \, ,
\;\; j \in \left[ 1,..., 2l+2\right]
\label{e1}\\
-i \, \left(
 \Xi \, \wedge \overrightharpoon{N}_{k-1,j}+(-)^{k}P \, \wedge \overrightharpoon{N}_{k+1,j}
\right)
& = &
\delta_{k,j} \, \mathrm{Id}_{(2l+2-j)} \, , \,
\nonumber \\
& &
j \in \left[ 1,..., 2l+2\right] \, , \, k \in \left[ 2,..., 2l+1\right]
\label{e2}\\
-i \,
\left(
 \Xi \, \wedge \overrightharpoon{N}_{2l+1,j}
\right)
&=&
\delta_{2l+2,j}. \mathrm{Id}_{(0)} \, ,
\;\; j \in \left[ 1,..., 2l+2\right]\, .
\label{e3}
\end{eqnarray}
A particular solution to the inhomogeneous eqns. (\ref{e1})-(\ref{e3}) on the diagonal $j = k$
is suggested by the Hodge decomposition of the Laplacian operator whose Fourier transform reads:
$\Xi \, \wedge \, P \, + \, P \, \wedge \, \Xi \, = \, p^{2} \, \mathrm{Id}$, and by the
identities $P \, \wedge \, P \, = 0$, $\Xi \, \wedge \, \Xi \, = 0$:
\begin{eqnarray}
\overrightharpoon{N}_{1,1}
& = &
 \; \frac{i}{p^{2}} \, ^{*}P_{(2l+1)}
\label{e4}\\
\overrightharpoon{N}_{j-1,j}
& = &
\; \frac{i}{p^{2}} \; P_{(2l+1-j)}, \;\; 2 \leq j \leq 2l+2
\label{e5}\\
\overrightharpoon{N}_{j+1,j}
& = &
- \; \frac{i}{p^{2}} \; \Xi_{(2l+1+j)}, \;\; 1 \leq j \leq 2l+1
\label{e6}
\end{eqnarray}
and all the other $\overrightharpoon{N}_{i,j}$ vanishing.
The particular solution thus found for the Fourier transform
$\overrightharpoon{N}_{_{1 \, ,\, 1}}$ of the propagator $<A^{(2l+1)} \otimes A^{(2l+1)}>$
involved in the computation of  Wilson $(2l+1)$-loops correlators turns out to be the so-called
Moore-Penrose pseudo-inverse\footnote{This can be most simply and explicitly checked in the three
dimensional case.
The projector $\Pi$ is then the projector transverse to $p$, which indeed corresponds to the
subspace of Fourier modes $\widehat{A}(p)$ such that $p^{\mu}\widehat{A}_{\mu}(p) = 0$ {\em i.e.} the
Fourier dual of the covariant gauge fixing condition $d^{*} A = 0$ imposed in $x$-space.}
of the operator $i\, ^{*}P$ which satisfies:
\begin{eqnarray}\label{pseudinv}
-i \, ^{*}P \, \overrightharpoon{N}_{_{1 \, , \, 1}}
& = &
\Pi
\label{eqnarray}
\end{eqnarray}
where $\Pi$ is the projector onto the subspace selected by the covariant gauge fixing condition.

\vspace{0.3cm}

\noindent
The propagators $<{\cal A}_{2l+2-j} \otimes {\cal A}_{2l+2-k}>$ might differ
from the particular solution above by terms corresponding to general solutions of the homogeneous
equations associated with eqns. (\ref{e1}) - (\ref{e3}) {\em i.e.} with all right hand sides vanishing.
The general solutions of these homogeneous equations on the space of tempered currents can be proven to
be forms with harmonic coefficients. Hence in the present case on $\mathbb{R}^{4l+3}$ with
Euclidean metrics the coefficient functions of these harmonic forms are harmonic polynomials of $(x-y)$.
In a first step we shall ignore such potential terms and consider the
$\overrightharpoon{N}_{jk}$ entirely given by eqns.(\ref{e4}) - (\ref{e6}).
We will comment on them in paragraph \ref{harm} and prove that they do not contribute insofar as we are only concerned with the computation of correlators of $(2l+1)$-loops.

\vspace{0.3cm}

\noindent
Performing the inverse Fourier transforms of  eqns.(\ref{e4}) - (\ref{e6}) yields the
explicit expressions of the $<{\cal A}_{j}(x){\cal A}_{k}(y)>$. The only one explicitly needed in the following is:
\begin{eqnarray}\label{5114}
\lefteqn{\left<
 {\cal A}^{(2l+1)}_{\mu_{1}, \cdots,\mu_{2l+1}}(x) \,
 {\cal A}^{(2l+1)}_{\nu_{1}, \cdots,\nu_{2l+1}}(y)
\right> }
\nonumber\\
& = &
 \; \frac{\Gamma \left( \frac{4l+3}{2} \right) }{2 \pi^{\frac{4l+3}{2}}}
\epsilon_{\mu_{1}, \cdots,\mu_{2l+1},\nu_{1}, \cdots,\nu_{2l+1},\rho} \,
\frac{(x-y)^{\rho}}{\left| x - y \right| ^{4l+3}}\, ,
\label{e7}
\end{eqnarray}
$\Gamma(w)$ being the Euler Gamma function and $\epsilon$ the $(4l+3)$-dimensional Levi-Civita symbol. The derivation of identity (\ref{5114}) relies on eqn. (\ref{propaTF}) of Appendix A.

\vspace{0.3cm}

\noindent
The gauge field theory is provided by the generating functional in presence of arbitrary source
currents $\vec{\cal J}$, which may be formally expressed by the following functional integral:
\begin{eqnarray}
{\cal Z}(\vec{\cal J})
& = &
{\cal N} \, \int {\cal D} {\vec A} \,
e^{2 i \pi k \, \left( \vec{\cal A} , D  \vec{\cal A} \right) \, +  \,
i \left( \vec{\cal A}, \vec{\cal J} \right)}
\label{functgen}
\end{eqnarray}
in which ${\cal D} {\vec A} \, \exp\{2 i \pi k \, ( \vec{\cal A} , D  \vec{\cal A} )\}$ is a
functional integration measure on some (unspecified) appropriate functional space.
This measure is assumed to have
all nice properties of usual gaussian integrals, and
${\cal N}$ is a normalization constant such that ${\cal Z}(\vec{\cal J}=0) = 1$.
The correlator of two $(2l+1)$-loops $\gamma_{1}$ and $\gamma_{2}$ is provided by the
quantity
\begin{eqnarray}
{\cal N} \, \int  {\cal D} {\vec A} \;
e^{2 i \pi k \, \left( \vec{\cal A} , D  \vec{\cal A} \right)}
\, e^{2 i \pi \, e_{1} \, \int_{\gamma_{1}}  A^{(2l+1)}}
\, e^{2 i \pi \, e_{2} \, \int_{\gamma_{2}}  A^{(2l+1)}}\, .
\label{link-tc1}
\end{eqnarray}
Let us represent the $(2l+1)$-loop $\gamma_{s}$ by the $(2l+2)$-current $j_{s}^{(2l+2)}$ so that
\begin{equation}\label{repres}
\int_{\gamma_{s}}  A^{(2l+1)} =
\int_{\mathbb{R}^{4l+3}}  A^{(2l+1)} \wedge  j_{s}^{(2l+2)}
\end{equation}
hence
\begin{equation}\label{repres2}
2 \pi \, e_{1} \, \int_{\gamma_{1}}  A^{(2l+1)} \, + \,
2 \pi \, e_{2} \, \int_{\gamma_{2}}  A^{(2l+1)}
=
\left( \vec{\cal A} , \vec{\cal J} \right)
\end{equation}
so that the loop correlator (\ref{link-tc1}) is given by eqn. (\ref{functgen}) identifying
\begin{equation}\label{identif}
\vec{\cal J}
=
2 \pi \left( e_{1} \, ^{*}j_{1}^{(2l+2)} + e_{2} \, ^{*}j_{2}^{(2l+2)},0,0, \cdots, 0 \right)\, .
\end{equation}
The phase in the integrand of eqn. (\ref{link-tc1}) involves:
\begin{equation}
k \, \left( \vec{\cal A}, D \vec{\cal A} \right) +
e_{1} \, \int_{\gamma_{1}}  A^{(2l+1)} +
e_{2} \, \int_{\gamma_{2}}  A^{(2l+1)}
=
k \left( \vec{\cal A}^{\prime}, D \vec{\cal A}^{\prime} \right)
-
\frac{1}{16 \pi^2 k} \, \left( \vec{\cal J}, D^{-1} \vec{\cal J} \right)
\label{link-tc2}
\end{equation}
where
\begin{eqnarray}\label{shift-sur-A}
\vec{\cal A}^{\prime} & = & \vec{\cal A} \, + \, \frac{1}{4 \pi k} \,  D^{-1} \vec{\cal J}\, .
\end{eqnarray}

\vspace{0.3cm}

\noindent
The functional space $\{\vec{\cal A}\}$ is assumed to be
stable\footnote{By passing let us notice that
any current $j^{(2l+2)}$ representing a $(2l+1)$-loop is such that $j^{(2l+2)} = d \eta^{(2l+1)}$,
the corresponding $^{*}j^{(2l+2)}$ thus belongs to the functional subspace of $\{A^{(2l+1)}\}$ obeying
the covariant gauge fixing condition $d^{*}A^{(2l+1)} =0$. Furthermore
this subspace is stable under the action of the operator $[D^{-1}]$ , cf. eqn. (\ref{pseudinv}),
so that this subspace is itself stable under the shift (\ref{shift-sur-A}).} under the shift
(\ref{shift-sur-A}). This shift is namely the counterpart of the one performed in eqn. (\ref{trivshift}),
and the gaussian properties of the functional measure
${\cal D} {\vec A} \, \exp\{2 i \pi k \, ( \vec{\cal A} , D  \vec{\cal A} )\}$ are the mere
counterparts of the Cameron-Martin property (\ref{21quart}). We thus proceed as in the
geometric approach.

\vspace{0.3cm}

\noindent
The functional integration leads to:
\begin{eqnarray}
{\cal N} \, \int  {\cal D} {\vec A} \;
e^{2 i \pi k \, \left( \vec{\cal A}, D \vec{\cal A} \right) }
\, e^{2 i \pi \, e_{1} \, \int_{\gamma_{1}}  A^{(2l+1)}}
\, e^{2 i \pi \, e_{2} \, \int_{\gamma_{2}}  A^{(2l+1)}}
& = &
e^{- \frac{i}{8 \pi k} \left( \vec{\cal{J}}, D^{-1} \vec{\cal{J}} \right) }\, .
\label{link-tc2b}
\end{eqnarray}

\noindent In the integral in the exponential in the r.h.s. of eqn. (\ref{link-tc2b}), the term of degree $(2l+1)$ is made of:
\bea
(D^{-1} \vec{\cal{J}}_{2l+1})_{\mu_{1}, \cdots,\mu_{2l+1}}(x) = \int_{\mathbb{R}^{4l+3}_{y}} \left< A^{(2l+1)}_{\mu_{1}, \cdots,\mu_{2l+1}}(x) A^{(2l+1)}_{\nu_{1}, \cdots,\nu_{2l+1}}(y) \right> \nonumber\\
(\vec{\cal{J}}_{2l+1})_{2l+1}^{\nu_{1}, \cdots,\nu_{2l+1}} d^{4l+3}y
\label{D^-1J}
\eea

\noindent and:
\bea
(\vec{\cal{J}}_{2l+1},D^{-1} \vec{\cal{J}}_{2l+1}) =  \int_{\mathbb{R}^{4l+3}_{x}} d^{4l+3}x (\vec{\cal{J}}_{2l+1})^{\mu_{1}, \cdots,\mu_{2l+1}}(x) (D^{-1} \vec{\cal{J}}_{2l+1})_{\mu_{1}, \cdots,\mu_{2l+1}}(x)\, .
\label{JD^-1J}
\eea

\noindent This yields two sorts of terms.
\begin{enumerate}
\item
Those of the form:
\begin{eqnarray}\label{link-ft4}
L(\gamma_{1},\gamma_{2})&\equiv& \int_{\mathbb{R}^{4l+3}_{x}} d^{4l+3}x ( ^{*}j_{1}^{(2l+2)})^{\mu_{1}, \cdots,\mu_{2l+1}}(x) (D^{-1} {^{*}j_{2}^{(2l+2)}})_{\mu_{1}, \cdots,\mu_{2l+1}}(x) \nonumber \\
& = &
 \int_{\mathbb{R}^{4l+3}_{x} \mbox{\footnotesize x} \mathbb{R}^{4l+3}_{y}}
 j_{1}^{(2l+2)}(x) \wedge \left< A^{(2l+1)}(x) \otimes A^{(2l+1)}(y) \right> \wedge j_{2}^{(2l+2)}(y) \nonumber \\
& = &
\frac{1}{(2l+1)!^2} \oint_{\gamma_{1}}
(dx^{\mu_{1}} \wedge \cdots \wedge dx^{\mu_{2l+1}}) \times
\,
\nonumber\\
&&
\;\;\;\;\;\;\;\;\;
\oint_{\gamma_{2}} (dy^{\nu_{1}} \wedge \cdots \wedge dy^{\nu_{2l+1}}) \left<
 A^{(2l+1)}_{\mu_{1}, \cdots,\mu_{2l+1}}(x) A^{(2l+1)}_{\nu_{1}, \cdots,\nu_{2l+1}}(y)
\right> \, .
\end{eqnarray}

\noindent They turn out to be the linking of $\gamma_{1}$ and $\gamma_{2}$ since after injecting expression (\ref{e7}) in the last line of eqn. (\ref{link-ft4}) one recognizes the generalized Gauss formula \cite{GCSS}.
The latter is recalled in Appendix B providing a consistency check of all normalizations between the
geometric and the ``naive" approaches. However, at variance with the virtue of the geometric approach,
it is important to notice in this respect that the values of the level $k$ and of the loop charges
$e_{j}$ are {\em not} quantized in the naive approach: their prescribed integer natures here are
{\em ad hoc} and imposed ``by hand".

This derivation sheds some light on the relation between the generalized Gauss formula
(\ref{link-ft4}) and the geometric approach developed in section \ref{4}. With respect to the variable $\vec{\cal{J}}$ the propagator identifies with $[{^{*}d}] _{MP}^{-1}$, the (Moore-Penrose pseudo-) inverse of ${^{*}d}$, whereas  it identifies with $[ d ]_{MP}^{-1}$ the inverse of $d$ with respect to the loops currents $j_{1}^{(2l+2)}$ and $j_{2}^{(2l+2)}$ in the following way. All loops are contractible in $\mathbb{R}^{4l+3}$, therefore there exists a de Rham current $\eta_{2}^{(2l+1)}$ such that:
\bea
j_{2}^{(2l+2)} =d \eta_{2}^{(2l+1)}\, ,
\eea

\noindent whose general solution is
\bea
\eta_{2}^{(2l+1)} = [ d ]_{MP}^{-1} j_{2}^{(2l+2)} + \zeta_{2}^{(2l+1)}\, ,
\eea

\noindent where $\zeta_{2}^{(2l+1)}$ is an arbitrary closed current. Indeed the current $\eta_{2}^{(2l+1)}$ is not unique since:
\bea
d (\eta_{2}^{(2l+1)} + \zeta_{2}^{(2l+1)}) = j_{2}^{(2l+2)}\, .
\eea

\noindent This reminds us of the definition of the Poincaré Homotopy:
\bea
\kappa \, \wedge \, d \, + \, d \, \wedge \, \kappa \, = \, \mathrm{Id}_{(2l+1)}
\eea

\noindent that encodes Poincaré Lemma (for $\mathbb{R}^{4l+3}$).
The degeneracy associated with the inversion of $d$ is exactly the one due to gauge
invariance since on $\mathbb{R}^{4l+3}$, and still by virtue of Poincar\'e's lemma, one has:
\[
\zeta_{2}^{(2l+1)}  \in Ker [d]
\Leftrightarrow
\exists \xi^{(2l+1)}, \; \zeta_{2}^{(2l+1)}  = d \, \xi^{(2l+1)}\, .
\]
We shall come back to this comment below when addressing the corresponding issue on
topologically non trivial $(4l+3)$-dimensional manifolds instead of $\mathbb{R}^{4l+3}$.

\item
It also involves the self-linkings of $(2l+1)$-loop $\gamma_{1}$ and of $(2l+1)$-loop $\gamma_{2}$
by means of formulas very similar to eqn. (\ref{link-ft4}), yet the integrals involved here are
ill-defined \cite{SLF,Po,JHW}.
An extrinsic procedure is required to have them make sense as quantities
defined modulo integers.
Framing provides {\em one} such procedure in the present case, a given integer
for each self-linking corresponding to a given framing choice.
By contrast the zero regularization
implemented in the geometric approach is less detailed as it does not prescribe any definite
integer value to any given self-linking.
\end{enumerate}

\subsubsection{Harmonic terms do not contribute}\label{harm}

So far we have ignored the presence of a harmonic contribution $H(x-y)$ to the propagator $<A^{(2l+3)}(x) \otimes A^{(2l+3)}(y)>$.
At first sight one might be tempted to argue that the absence of such terms is implied by the cluster property meaning that $<A^{(2l+3)}(x) \otimes A^{(2l+3)}(y)> \to 0$ when $||x-y|| \to + \infty$. However this is i) beside the point ii) not necessarily true.

\vspace{0.3cm}

\noindent i) It is beside the point insofar as we are interested in correlators of $(2l+1)$-loops {\it i.e. closed} curves. Assuming that the propagator involves such a harmonic term $H(x-y)$, let us
generalize eqn. (\ref{link-ft4}) by
\begin{eqnarray}\label{link-ft4bis}
\lefteqn{ \widetilde{L}(\gamma_{1},\gamma_{2})} \nonumber \\
&=& \int_{\mathbb{R}^{4l+3}_{x} \mbox{\footnotesize x} \mathbb{R}^{4l+3}_{y}}
 j_{1}^{(2l+2)}(x) \wedge \left\{\left< A^{(2l+1)}(x) \otimes A^{(2l+1)}(y) \right> + H(x-y) \right\} \wedge j_{2}^{(2l+2)}(y) \nonumber \\
& \equiv &
L(\gamma_{1},\gamma_{2}) + L'_H(\gamma_{1},\gamma_{2})
\end{eqnarray}

\noindent The currents $j_{1,2}^{(2l+2)}$ dualize $(2l+1)$-loops so that \textit{e.g.} $j_{1}^{(2l+2)} = d \eta_{1}^{(2l+1)}$ so that through integration by part,
\begin{eqnarray}\label{link-ft4ter}
L'_H(\gamma_{1},\gamma_{2}) & = &
\int_{\mathbb{R}^{4l+3}_{x} \mbox{\footnotesize x} \mathbb{R}^{4l+3}_{y}} \eta_{1}^{(2l+1)}(x) \wedge \left( d_y H(x-y) \right) \wedge j_{2}^{(2l+2)}(y) \nonumber \\
& = & 0
\end{eqnarray}

\noindent This suggests that the appropriate functional space on which the propagator has to be defined is a quotient modulo harmonic parts. Such a functional space has been studied in ref. \cite{Tr}.

\vspace{0.3cm}

\noindent By passing, eqn. (\ref{link-ft4ter}) proves that harmonic contributions vanish even when $j_{2}^{(2l+2)}$ dualizes a non compactly supported loop, such as a $(2l+1)$-hyperplane. This property is expected to be particularly relevant in order to extend the present result to the sphere $S^{4l+3}$.

\vspace{0.3cm}

\noindent ii) The cluster property may not hold with another gauge fixing choice. See for instance the 3-dimensional case with axial gauge fixing.

\subsubsection{Impact of the gauge fixing choice}

Equation (\ref{link-ft4}) was noticed to reproduce the generalized Gauss formula when the propagator
$<A^{(2l+3)} \otimes A^{(2l+3)}>$ is given by eqn. (\ref{e7}). Another condition than the gauge fixing (\ref{gauge-fix-3d}) would lead to a different propagator. Equation (\ref{link-ft4}) would
then provide an expression of the linking number different from the one obtained using the generalized Gauss invariant. For example in the three dimensional case, the ``axial gauge" choice leads to a braiding interpretation of the linking number \cite{DBSS}, rather than the solid angle interpretation reminded in Appendix B.
Let us stress that all gauge fixing choices are equivalent ways of computing the generalized linking number. Indeed, the propagator in the covariant gauge and one with an alternative gauge choice differ by terms involving the derivative $d$ whose actions on the closed currents dualizing $(2l+1)$-loops vanish. In a Quantum Electro-Dynamical language, the latter are ``conserved
currents" which guarantees the gauge fixing independence of observables associated with these
currents.

\subsection{Further issues arising on the ${\cal S}^{4l+3}$ then on further non trivial
manifolds}

As we already mentioned it, Chern-Simons field theory cannot provide a quantization of the
level $k$ nor of the charge $q$. This is due to the fact such a theory is developed over the non
compact space $\mathbb{R}^{4l+3}$. It's only when going on a closed manifold such as a sphere that the quantization naturally appeared in the geometric approach. This suggest that to get such a
quantization of $k$ and $q$ within the field theoretic framework, one should have to first define
a field theory over a closed manifold $M$, starting with $S^{4l+3}$. Since the CS lagrangian is not a globally defined 3-form, we anticipate two possible paths: one based on a partition of unity subordinated to a good covering of $M$ and a second based on a polyhedral decomposition of $M$.

\begin{enumerate}

\item
We could consider a polyhedral decomposition $\Delta$ of $M$ and start with field theories on each of the fundamental \textit{i.e.} $(4l+3)$-dimensional polyhedra $\Delta_{\alpha}$ of the decomposition. Once this done on fundamental polyhedra we would have to see how things match on the $(4l+2)$-dimensional boundaries $\Delta_{\alpha \beta}$ of these polyhedra leading to $(4l+2)$-dimensional field theories on those boundaries. We would have to keep proceeding along this line till we reach the polyhedral elements of dimension $0$ of the decomposition. This would be related to the short formula defining the integral of a DB class, as explained in \cite{BGST}.

\item
We could provide $M$ with a partition of unity subordinated to a good covering ${\cal U}$ in such a way that each open set ${\cal U}_\alpha$ supports a field theory in $\mathbb{R}^{4l+3}$. Matching these theories in the $(4l+3)$-dimensional intersections ${\cal U}_{\alpha \beta}$ would lead to considering extra field theories in these intersections then in the triple intersections ${\cal U}_{\alpha \beta \gamma}$ etc. The present point of view in which all supplemented field theories would be on $\mathbb{R}^{4l+3}$ is a smoothing of the former polyhedral approach. This would be related to the long formula appearing in \cite{BGST}.

\end{enumerate}

We would like to stress out that our procedure to compute the propagator of the abelian CS field theory on $\mathbb{R}^{4l+3}$  exhibits a set of descent equations whose resolution is made simple because $\mathbb{R}^{4l+3}$ has no cohomology (except in dimension $0$). Our results might be extended to $S^{4l+3}$ since it shares the same cohomology properties for the concerned degrees. In the case of a general closed manifold, such has $S^{2l+1} \times S^{2l+2}$, this would not be true. However, locally that is to say with respect to a good covering and with an Euclidean metric on each open set, such a descent might still hold. Yet the gluing constraints on the whole manifold (\textit{e.g.} via a partition of unity) would prevent the descent from being globally trivial. The simplest case to investigate would be $S^3$ and the first non trivial one $S^1 \times S^2$.

\vspace{0.3cm}

\noindent
Concerning the propagator itself,
the fact it coincides with the Gauss integral is
once more only due to the fact we are working on $\mathbb{R}^{4l+3}$. One would expect a
different expression for the propagator on a closed manifold. However there exist
expressions of the Gauss integral on spheres \cite{DTG}. One could also try to mimic Gauss
zodiacus idea, at least in the case of $S^3$ identified with $SU(2)$, replacing the notion of
translations acting on $\mathbb{R}^{3}$ by actions on $SU(2)$. From the point of view of the two
possible approaches previously mentioned, we can expect a collection of propagators,
associated with the different field theory arising from the construction (for instance one for each polyhedra type of the decomposition of the closed manifold), but also a gluing rule explaining how these propagators "communicate".

It appears as a very interesting problem how this could be properly handled because it would provide
an example of a field theory over a closed manifold. We can have some hope about how this can be done, because
the theory which we are dealing with is a topological one, and also because the geometric
approach provides us with the final answer concerning Wilson observables.

\section{Conclusions and outlook}

The treatment of abelian Chern-Simons to generate link invariants introduced in \cite{GT}
straightforwardly extends to the case of oriented closed $(4l+3)$-dimensional manifolds without torsion. Actually, we didn't show that the expectation values of our generalised Wilson lines are ambient isotopy invariants. This can be easily checked extending what has been done in \cite{GT}. In the same way, it is possible to establish satellite relations for our generalised invariants. As for torsion, one could follow the approach developed for  $\mathbb{R}P^3$ in \cite{Th}. One can wonder whether the DB strategy applies more generally to abelian BF systems. Using Deligne-Beilinson Cohomology technics might also provide a way to study higher order systems, that is to say systems whose classical lagrangian involves DB products of more than two DB classes. In any of these cases one should expect homology and intersection to play the fundamental role.

\newpage

\subsection*{Appendix A: Forms and Fourier Transform}

This appendix is devoted to the conventions and properties of Fourier transform applied to
forms and linear operators acting on them. These properties are used in Section \ref{sect5}
in order to evaluate precisely the propagator of the vector potential in the covariant gauge.

\subsubsection*{Berezin-Fourier transform preserving forms degrees}

The components of a $q$-form are defined through
\begin{equation}
B^{(q)}= B(x)_{\nu_1 ... \nu_q} \, \psi^{\nu_1}\wedge ... \wedge \psi^{\nu_q}
\end{equation}
where $\psi^{\mu}= dx^{\mu}$. This convention partially avoids clutter with factorial numbers.

\noindent
The Fourier transform of a $q$-form is then defined as
\begin{eqnarray}
\overrightharpoon{B}^{(q)}
&\equiv&
\left[\int d^{n}x \, \mathrm{e}^{ip_{\mu}x^{\mu}} B(x)_{\nu_1 ... \nu_q}\right]
\left[\frac{1}{l_{(n-q)}} \int d^{n}\psi \,
\mathrm{e}^{i\bar{\omega}_{\mu}\psi^{\mu}} \psi^{\nu_1}\wedge ... \wedge \psi^{\nu_q} \right]
\nonumber \\
&=&
\left[\int d^{n}x \, \mathrm{e}^{ip_{\mu}x^{\mu}} B(x)_{\nu_1 ... \nu_q}\right]
\nonumber \\
&&
\frac{\epsilon^{\nu_{q+1}...\nu_n...\nu_1...\nu_q}}{(n-q)!}
\frac{\epsilon^{\tau_{q+1}...\tau_n...\mu_1...\mu_q}}{q!}
\delta_{\nu_{q+1}\tau_{q+1}} \, ... \,
\delta_{\nu_{n}\tau_{n}} \, \bar{\omega}_{\mu_1}\wedge ... \wedge \bar{\omega}_{\mu_q}
\nonumber \\
&=&
\frac{\wideparen{B}(p)_{\nu_1...\nu_q}}{q!(n-q)!} \,
\epsilon^{\nu_{q+1}...\nu_n...\nu_1...\nu_q} \,
\delta_{\nu_{q+1}\tau_{q+1}} \, ... \,
\delta_{\nu_{n}\tau_{n}} \, \epsilon^{\tau_{q+1}...\tau_n...\mu_1...\mu_q} \,
\bar{\omega}_{\mu_1}\wedge \, ... \, \wedge \bar{\omega}_{\mu_q}
\nonumber \\
&=&
\wideparen{B}(p)_{\nu_1...\nu_q} \,
\delta^{\nu_{1}\mu_{1}} \, ... \, \delta^{\nu_{q}\mu_{q}} \,
\bar{\omega}_{\mu_1}\wedge \, ... \, \wedge \bar{\omega}_{\mu_q}
\nonumber \\
&=&
\wideparen{B}(p)^{\mu_1 ... \mu_q} \,
\bar{\omega}_{\mu_1}\wedge \, ... \,  \wedge \bar{\omega}_{\mu_q}
\end{eqnarray}
where $l_{(a)}=1$ if $a$ is even and $l_{(a)}=i$ if $a$ is odd, $\bar{\omega}_{\mu} \equiv dp_{\mu}$, and  $\, \wideparen{} \,$  denotes the usual Fourier transform on functions. With this definition, the Fourier transform of a  $q$-form is itself a  $q$-form, that is to say the Fourier transform respects the form degrees.

\noindent
Inverse Fourier transform is accordingly defined as
\begin{eqnarray}
\overleftharpoon{B}^{(q)} &\equiv& \frac{1}{(2\pi)^n}
\left[\int d^{n}p \, \mathrm{e}^{-ip_{\mu}x^{\mu}} B(p)^{\nu_1 ... \nu_q}\right]
\left[\frac{1}{l_{(n-q)}}
\int d^{n}\bar{\omega} \,
\mathrm{e}^{-i\bar{\omega}_{\mu}\psi^{\mu}}
\bar{\omega}_{\nu_1}\wedge \, ... \, \wedge \bar{\omega}_{\nu_q} \right]
\nonumber \\
&=&
\frac{\widecheck{B}(x)^{\nu_1...\nu_q}}{q!(n-q)!}
\epsilon_{\nu_{q+1}...\nu_n...\nu_1...\nu_q} \,
\delta^{\nu_{q+1}\tau_{q+1}} \, ... \,
\delta^{\nu_{n}\tau_{n}} \, \epsilon_{\tau_{q+1}...\tau_n...\mu_1...\mu_q} \,
\psi^{\mu_1}\wedge \, ... \, \wedge \psi^{\mu_q}
\nonumber \\
&=&
\widecheck{B}(x)^{\nu_1...\nu_q} \,
\delta_{\nu_{1}\mu_{1}} \, ... \, \delta_{\nu_{q}\mu_{q}} \,
\psi^{\mu_1}\wedge \, ... \, \wedge \psi^{\mu_q}
\nonumber \\
&=&
\widecheck{B}(x)_{\mu_1 ... \mu_q} \,
\psi^{\mu_1}\wedge \, ... \, \wedge \psi^{\mu_q}
\end{eqnarray}

\noindent where $\, \widecheck{} \,$ is the inverse Fourier transform on functions. An explicit evaluation indeed confirms that

\begin{equation}
\overleftharpoon{\overrightharpoon{B}} \; = \, B\, .
\end{equation}

\noindent An important property is that the Hodge operation and Berezin-Fourier transform do commute:
\begin{eqnarray}
\lefteqn{ ^{*} \! \left[\frac{1}{l_{(n-q)}}\int d^{n}\psi \,
\mathrm{e}^{i\bar{\omega}_{\mu}\psi^{\mu}}
\psi^{\nu_1}\wedge \, ... \, \wedge \psi^{\nu_q}  \right] }
\nonumber \\
&=&
\frac{1}{q!} \,
\delta^{\nu_1 ... \nu_q}_{\sigma_1 ... \sigma_q} \,
\delta^{\sigma_1 \mu_1} \, ... \, \delta^{\sigma_q \mu_q} \,
\bar{\omega}_{\mu_1}\wedge \, ... \, \wedge \bar{\omega}_{\mu_q}
\nonumber \\
&=&
\left[\frac{1}{l_{(n-q)}}\int d^{n}\psi \, \mathrm{e}^{i\bar{\omega}_{\mu}\psi^{\mu}}
{} ^{*} \! \left( \psi^{\nu_1}\wedge \, ... \, \wedge \psi^{\nu_q}\right) \right]\, .
\end{eqnarray}

\subsubsection*{An useful Fourier transform}

The explicit computation of the fundamental propagator (\ref{e7}) relies on the following Fourier transform
\begin{equation}\label{propaTF}
\overleftharpoon{(\frac{p^{\tau}}{p^2})}\, =\,  \frac{1}{(2\pi)^{4l+3}}\int d^{4l+3}p\, \mathrm{e}^{-ip_{\mu}x^{\mu}} \, \frac{p^{\tau}}{p^2} \, = \, -i \frac{\Gamma \left( \frac{4l+3}{2} \right) }{2 \pi^{\frac{4l+3}{2}}}\frac{x^{\tau}}{x^{4l+3}} \, .
\end{equation}

\subsubsection*{Berezin-Fourier transform for linear operators}
The Berezin-Fourier transform of a linear operator ${\cal O}$ acting on forms is defined by
\begin{equation}
\overrightharpoon{\cal O}B \equiv \, \overrightharpoon{{\cal O} \overleftharpoon{B}} \, .
\end{equation}

\noindent Accordingly, the (useful) Fourier transform of the differential, its Hodge dual and the co-differential read:
\begin{eqnarray}\label{PetXi}
\overrightharpoon{d} &=& -i p^{\mu}\bar{\omega}_{\mu} \equiv -i P \\
\overrightharpoon{(^{*}d)}&=&^{*} \!
\left( \overrightharpoon{d} \right) = -i \, ^{*} \! P \\
\overrightharpoon{\delta}&=& ^{*} \!
\left(\overrightharpoon{d}\right) \! {}^{*}= -i \, ^{*} \! P ^{*} \equiv -i \, \Xi\, .
\end{eqnarray}

\newpage

\subsection*{Appendix B: Generalized Gauss linking number}

\subsubsection*{Definition of the linking number}

We consider two  $(2l+1)$-dimensional closed surfaces $\gamma_{2l+1}$ and $\gamma_{2l+1}'$ embedded in the space $\mathbb{R}^{4l+3}$.
They are defined as a map from the $(2l+1)$-dimensional closed manifold $T$, respectively  $T'$, to $\mathbb{R}^{4l+3}$.
Their linking number is given by \cite{GCSS}
\begin{equation}
L(\gamma_{2l+1},\gamma_{2l+1}')= {\cal N}_{l} \oint_{\gamma_{2l+1}}
dx^{\mu} \oint_{\gamma_{2l+1}'}dy^{\nu} \epsilon_{\mu,\nu,\sigma} \,
\delta^{\sigma \tau} \partial_{\tau}
\vert x-y \vert^{-4l-1}
\label{linknumGCSS}
\end{equation}
where the $x$s (resp. $y$s) are the coordinates of points of  $\gamma_{2l+1}$ (resp. $\gamma_{2l+1}'$) and $\epsilon$ is the $(4l+3)$-dimensional Levi-Civita symbol. We have used the following
shorthand notations
\begin{equation}
dx^{\mu}=dx^{\mu_1} \cdots dx^{\mu_{2l+1}}
 ~,~
dy^{\nu}=dy^{\nu_1} \cdots dy^{\nu_{2l+1}}
~,~
\epsilon_{\mu,\nu,\sigma}=\epsilon_{\mu_1 \cdots \mu_{2l+1}\nu_1 \cdots \nu_{2l+1}\sigma}
\end{equation}
and set $\partial_{\tau} = \partial_{y^{\tau}}$. The other choice of the derivative,  $\partial = \partial_x$,
reverses the sign of the linking number, e.g. it corresponds to an orientation choice.
The normalisation of the linking number is
\begin{equation}
{\cal N}_{l} = \frac{\Gamma \left( \frac{4l+3}{2}\right)}{(8l+2)\sqrt{\pi^{4l+3}}(2l+1)!^2} \, .
\end{equation}
with $\Gamma$ the Euler Gamma function, satisfying $\Gamma(n+1)=n!$ for an integer $n$.

The linking number can be given a more enlightening form as follows.
For two points $x$ (resp $y$) on  $\gamma_{2l+1}$ (resp. $\gamma_{2l+1}'$), we consider the unitary vector
\begin{equation}
e_{xy} = \frac{x-y}{\vert x-y \vert} .
\end{equation}
The unitary vector $e_{xy}$ thus defines a map from $T \times T'$ to the sphere $S^{4l+2}$
whose degree is the linking number \cite{vW}.
We now consider the quantity
\begin{equation}
[e_{xy} ; dx ; dy] = \frac{1}{(2l+1)!^2} \epsilon_{\mu,\nu, \sigma}\, dx^{\mu} dy^{\nu} e_{xy}^{\sigma}
\end{equation}
which has a simple physical interpretation:
\begin{equation}
\frac{[e_{xy} ; dx ; dy]}{\vert x-y \vert^{4l+2}}
\end{equation}
is the oriented solid angle formed by a simultaneous displacement $dx$ on $\gamma_{2l+1}$ and $dy$ on $\gamma_{2l+1}'$.

The linking number can thus be given the following equivalent form
\begin{equation}
L(\gamma_{2l+1},\gamma_{2l+1}')= \frac{1}{S_{4l+2}} \oint_{\gamma_{2l+1}} \oint_{\gamma_{2l+1}'}
\frac{[e_{xy} ; dx ; dy]}{\vert x-y \vert^{4l+2}}
\label{linknumD}
\end{equation}
and interpretation of a global solid angle.
We have used the value of the surface of a unit sphere $S^n$ is given by
\begin{equation}
S_n = \frac{2 \pi^{\frac{n+1}{2}}}{\Gamma (\frac{n+1}{2})} ~.
\end{equation}
This is also the total solid angle in dimension $n+1$.

\subsubsection*{The three dimensional case}

In the three dimensional case ($l=0$), the linking number (\ref{linknumD}) is the famous Gauss invariant \cite{GE}
\begin{equation}
L(\gamma,\gamma')= \frac{1}{4\pi} \oint_{\gamma}  \oint_{\gamma'}d\vec{x}\times d\vec{y}. \,
\frac{\vec{x}-\vec{y}}{\vert \vec{x}-\vec{y} \vert^3} \, .
\label{linknum3}
\end{equation}

The unitary vector
\begin{equation}
\vec{e}_{xy} = \frac{\vec{x}-\vec{y}}{\vert x-y \vert} \,  .
\end{equation}
defines a map $e$ from $S^1 \times S^{1}$ to the sphere $S^{2}$ whose degree is the linking number \cite{vW}.
The image of the map $e$ is generically a surface called the zodiacus by Gauss who also obtained a necessary
condition for a point to be on its boundary: the tangent vectors to the two curves at
points $x$ and $y$ respectively and the vector $\vec{e}_{xy}$ are linearly dependent.
In other words, these are points such that
\begin{equation}
[\vec{e}_{xy} ; d\vec{x} ; d\vec{y}]=0
\end{equation}
and do not contribute to the Gauss integral. This condition is only necessary and not all solutions do represent actual
boundaries of the zodiacus. Two cases have to be distinguished: (1) the two curves are not linked and the zodiacus has at least
one boundary, (2) the two curves are linked and the curve defined by the previous condition cannot be a boundary of the zodiacus
which is in fact the whole sphere.

Some intuition on these matters can be given by the following particular case. We consider a basic configuration of two
circles $\gamma$, having radius one and centered at the origin, and  $\gamma '$, having radius $R$ greater than one. This
configuration has linking number one when the circle $\gamma '$ intersects the disc defined by $\gamma$. In the extreme case
where the radius $R \rightarrow \infty$, the $\gamma '$ circle may be deformed to a straight
line perpendicular to the plane containing the circle $\gamma$ completed with an half circle at infinity whose contribution
to the Gauss integral vanishes.

The circle $\gamma$ can be parameterized as
\begin{equation}
x_1 = \mathrm{cos}(s) \, , \, x_2 = \mathrm{sin}(s)\,, \,x_3=0
\end{equation}
and the straight line $\gamma$' as
\begin{equation}
y_1=0\,, \,y_2=y
\end{equation}
and intersection with the disc bounded by $\gamma$ occurs when $\vert y \vert < 1$.

We obtain the linking number by integrating over the straight line
\begin{equation}
L(\gamma,\gamma')= \frac{1}{4\pi}
\int_{0}^{2\pi} ds \int_{-\infty}^{+\infty}dy_3\,  \frac{1 -y \, \mathrm{sin}(s)}{(1-2y\,\mathrm{sin}(s) +y^2 +y_3^2)^{\frac{3}{2}}}
\end{equation}
The integral over $y_3$ is classical and, for $\vert y \vert \neq 1$, one has
\begin{equation}
L(\gamma,\gamma')= \frac{1}{2\pi}
\int_{0}^{2\pi} ds \,  \frac{1 -y \, \mathrm{sin}(s)}{(1-2y\,\mathrm{sin}(s) +y^2)}
\end{equation}
The evaluation of this integral can be done by expanding the integrand in powers of the sine, using then the classical
values of integral of even powers of the sine function. The result is then
\begin{equation}
L(\gamma,\gamma')= 1 \,\, \mathrm{for} \,\, \vert y \vert < 1 \, , \, L(\gamma,\gamma')= 0 \,\, \mathrm{for} \,\,\vert y \vert > 1 \, .
\end{equation}

The unitary vector $\vec{e}$ reads
\begin{equation}
\vec{e} = \frac{\mathrm{cos}(s)\, \vec{i}+(\mathrm{sin}(s) - y)\vec{j}-y_3 \vec{k}}{(1-2y\, \mathrm{sin}(s) +y^2 +y_3^2)^{\frac{1}{2}}}
\end{equation}
and the necessary condition for a point to be on the boundary of the zodiacus is
\begin{equation}
1-y\, \mathrm{sin}(s)=0.
\end{equation}
A moment thought shows that for $\vert y \vert <1$, there is no boundary and the vector $\vec{e}$ sweeps the whole sphere once.
On the contrary, for $\vert y \vert >1$, the zodiacus has two boundaries at the values $s=\mathrm{arcsin}(y^{-1})$ and
$s=\pi -\mathrm{arcsin}(y^{-1})$ that join at antipodal points for $y_3=\pm \infty$.

\subsubsection*{Higher dimensional cases}

As in the three dimensional case, the unitary vector $e_{xy}$ spans on the sphere $S^{4l+2}$ the zodiacus
associated with the two surfaces $\gamma_{2l+1}$
and $\gamma_{2l+1}'$. The eventual boundaries of the zodiacus necessarily correspond to stationary points of $e_{xy}$
upon infinitesimal displacements $\delta x$ (resp. $\delta y$) on the surface $\gamma_{2l+1}$ (resp. $\gamma_{2l+1}'$),
that is to say $\delta e_{xy}=0$
where
\begin{equation}
\delta e_{xy} = \frac{\delta(x-y)- e_{xy}(e_{xy}.\delta(x-y))}{\vert x-y \vert}
\end{equation}
If the surfaces $\gamma_{2l+1}$ and $\gamma_{2l+1}'$ are parameterized by (even local) coordinates $s_i$, $t_j$
respectively ($i,j = 1 \,  ... \,2l+1$), then
\begin{equation}
\delta(x-y) = a_i \frac{\partial x}{\partial s_i} - b_j  \frac{\partial y}{\partial t_j}
\end{equation}
where $a_i$ and $b_j$ are two families of infinitesimal coefficients. As a consequence of the stationarity conditions,
the vector $e_{xy}$ is thus a linear combination of the $4l+2$ tangent vectors $\partial_{s_i}x$ and $\partial_{t_j}y$. Hence the oriented solid
angle formed by two simultaneous displacements on both curves vanishes at the boundary of the zodiacus:
\begin{equation}
[e_{xy} ; \partial_i x ; \partial_j y] =0.
\end{equation}

We shall now check the normalisation of the linking number considering a simple choice of linked surfaces.
We choose a $(2l+1)$-sphere centered at the origin and an orthogonal $(2l+1)$-hyperplane containing the origin.
They are given respectively by
\begin{equation}
 \gamma_{2l+1} : x_1^2+\cdots + x_{2l+2}^2=1, \,\, x_{2l+3}= \cdots = x_{4l+3}=0
\end{equation}
and a $(2l+1)$-hyperplane
\begin{equation}
 \gamma_{2l+1}': y_1 =\cdots = y_{2l+2}=0
\end{equation}
with its completion (an half-sphere) at infinity whose contribution to the Gauss integral vanishes. The ball defined
by the sphere $\gamma_{2l+1}$ and the hyperplane $\gamma_{2l+1}'$ intersect at the origin so we have a configuration with linking
number equal to one and a moment thought shows that the zodiacus is the whole $(4l+2)$-sphere.

The linking number (\ref{linknumD}) here reads
\begin{equation}
L(\gamma_{2l+1},\gamma_{2l+1}')= \frac{1}{S_{4l+2}}
\oint_{\gamma_{2l+1}} d^{2l+1}x
\oint_{\gamma_{2l+1}'}d^{2l+1}y \frac{1}{(1+\vert\vec{y}\vert^2)^{\frac{4l+3}{2}}} .
\end{equation}
The first integral yields the surface of the $(2l+1)$-sphere
\begin{equation}
\oint_{\gamma_{2l+1}} d^{2l+1}x =  S_{2l+1} \, ,
\end{equation}
while the second integral can be decomposed in a surfacic and a radial ones as
\begin{equation}
\oint_{\gamma_{2l+1}'}d^{2l+1}y \, \frac{1}{(1+\vec{y}^2)^{\frac{4l+3}{2}}}=S_{2l} \int_{0}^{\infty}dy \, \frac{y^{2l}}{(1+y^2)^{\frac{4l+3}{2}}}
\end{equation}
The radial integral is a classic one and may be computed after the change of variable $y = \mathrm{tan}(\theta)$
\begin{equation}
\int_{0}^{\infty}dy \, \frac{y^{2l}}{(1+y^2)^{\frac{4l+3}{2}}}
=
\int_{0}^{\frac{\pi}{2}}d\theta \, \mathrm{sin}^{2l}(\theta) \mathrm{cos}^{2l+1}(\theta)
=
\frac{\Gamma(l+\frac{1}{2})\Gamma (l+1)}{2\Gamma (2l+\frac{3}{2})}~.
\end{equation}

We thus obtain
\begin{equation}
L(\gamma_{2l+1},\gamma_{2l+1}')= \frac{S_{2l}S_{2l+1}}{S_{4l+2}}
\frac{\Gamma(l+\frac{1}{2})\Gamma (l+1)}{2\Gamma (2l+\frac{3}{2})}
\end{equation}
what drastically simplifies into the expected result
\begin{equation}
L(\gamma_{2l+1},\gamma_{2l+1}')= +1~.
\end{equation}

\newpage

\vskip 1 truecm

\vfill\eject

\end{document}